\documentclass[lettersize,journal]{IEEEtran}

\usepackage{amsmath,amsfonts}
\usepackage{array}
\usepackage[caption=false,font=normalsize,labelfont=sf,textfont=sf]{subfig}
\usepackage{textcomp}
\usepackage{stfloats}
\usepackage{url}
\usepackage{verbatim}
\usepackage{graphicx}
\usepackage{cite}

\usepackage{bbm}
\usepackage{amssymb}
\usepackage{caption}
\usepackage[short]{optidef}
\usepackage[pagebackref=true,breaklinks=true,letterpaper=true,colorlinks,bookmarks=false]{hyperref}
\usepackage{booktabs}
\usepackage{multirow}
\usepackage{algorithmicx}
\usepackage{algpseudocode}
\usepackage{algorithm}
\usepackage{adjustbox}
\usepackage{fancyhdr}

\renewcommand{\appendixname}{Supplementary material}
\newcommand\E{\mathbf E}
\newcommand\A{\mathbf A}

\newcommand\X{\mathbf X}
\newcommand\N{\mathbf N}
\newcommand\B{\mathbf B}

\newcommand\Real{\mathbb R}
\newcommand\1{\mathbbm 1}

\newcommand\abold{\mathbf a}
\newcommand\bbold{\mathbf b}
\newcommand\ubold{\mathbf u}
\newcommand\zbf{\mathbf z}
\newcommand\xbf{\mathbf x}
\newcommand\ebf{\mathbf e}

\newcommand\ie{{\it{i.e. }}}
\newcommand\etal{{\it{et al. }}}

\title{Entropic Descent Archetypal Analysis\\for Blind Hyperspectral Unmixing}

\author{Alexandre Zouaoui\thanks{Alexandre Zouaoui, Jocelyn Chanussot, and Julien Mairal are with Univ. Grenoble Alpes, Inria, CNRS, Grenoble
INP, LJK, 38000 Grenoble, France. Gedeon Muhawenayo worked on this project while he was there.}, 
Gedeon Muhawenayo, Behnood Rasti\thanks{Behnood Rasti is with Helmholtz-Zentrum Dresden-Rossendorf (HZDR)},~\IEEEmembership{Senior Member,~IEEE}  \\ Jocelyn Chanussot,~\IEEEmembership{Fellow,~IEEE,} and Julien Mairal,~\IEEEmembership{Senior Member,~IEEE} 
}

\fancypagestyle{azouaoui}{%
  \fancyhf{} % clear all fields
  
  \fancyhead[C]{This work has been submitted to the IEEE for possible
    publication.
    }
}%

\makeatletter
\let\ps@IEEEtitlepagestyle\ps@azouaoui
\makeatother

\begin{document}

\maketitle

\begin{abstract}
In this paper, we introduce a new algorithm based on archetypal analysis 
for blind hyperspectral unmixing, 
assuming linear mixing of endmembers. Archetypal analysis is a natural
formulation for this task. This method does not require the presence
of pure pixels (i.e., pixels containing a single material) but instead represents endmembers as convex combinations of a few pixels present in the
original hyperspectral image. 
Our approach leverages an entropic gradient descent strategy, which (i) provides
better solutions for hyperspectral unmixing than traditional archetypal analysis algorithms, and (ii) leads to efficient GPU implementations.
Since running a single instance of our algorithm is fast, we also propose
an ensembling mechanism along with an appropriate model selection
procedure
that make our method robust to hyper-parameter choices
while keeping the computational complexity reasonable.
By using six standard real datasets, we show that our approach outperforms
state-of-the-art matrix factorization and recent deep learning methods.
We also provide an open-source PyTorch implementation: \href{https://github.com/inria-thoth/EDAA}{https://github.com/inria-thoth/EDAA}.
\end{abstract}


\section{Introduction}

\label{sec:intro}

Hyperspectral imaging (HSI) \cite{landgrebe_hyperspectral_2002,
  plaza_recent_2009, schaepman_earth_2009,goetz_imaging_1985, green_imaging_1998}
consists of measuring the electromagnetic spectrum in a scene by using multiple narrow spectral bands.
Thanks to its richer spectral information compared to traditional RGB images,
HSI enables more accurate materials identification, leading to a
broad range of applications including crop monitoring in agriculture
\cite{adao_hyperspectral_2017}, waste sorting \cite{karaca_automatic_2013},
food safety inspection \cite{gowen_hyperspectral_2007}, or mineralogy~\cite{fox_applications_2017}.

Remote sensing \cite{clark_imaging_2003, bioucas-dias_hyperspectral_2013}, such as airborne or satellite imagery, yields hyperspectral
images whose pixels capture several objects or materials.
As such, each pixel can include several pure spectral components (called \emph{endmembers}), mixed in
different proportions \cite{ghamisi_advances_2017}.
Any further analysis hence requires identifying and disentangling endmembers present in a scene before estimating their respective proportions, or
fractional \emph{abundances},
within each pixel of the HSI \cite{parra_unmixing_1999}.
Since the endmembers spectrum signatures are not known beforehand and must be
estimated from data, this
operation is named \emph{blind} hyperspectral unmixing (HU)
\cite{keshava_spectral_2002, bioucas-dias_hyperspectral_2012} owing to its
link with blind source separation \cite{comon_handbook_2010}.

In this paper, we adopt a linear mixing model since it is often relevant
in remote sensing scenes where mixtures occur between macroscopic materials.
Therefore, we assume that each observed pixel can be represented as a
linear combination of endmembers and some additive noise.
In other words, we are interested in tackling unsupervised linear HU \cite{parra_unmixing_1999}.

Further assumptions on the nature of endmembers are generally needed to
estimate meaningful spectra.
For instance, it can be assumed that there exists at least one pure pixel
for each material present in the scene.
The problem then requires finding these pure pixels within the original image.
The pure pixel assumption is at the core of several geometrical endmember extraction
methods including pixel purity index (PPI) \cite{boardman_mapping_1995}, N-FINDR
\cite{winter_n-findr_1999} and vertex component analysis (VCA)
\cite{nascimento_vertex_2005}.
Once endmembers have been extracted, abundances can be estimated by
minimizing the least squares errors between the original input spectra and the
linearly reconstructed spectra as long as the abundances fractions satisfy the
two physical constraints stating that they should be non-negative and sum
to one for each pixel \cite{heinz_fully_2001}. 
That being said, pure pixels are often missing in real scenarios.
In the absence of pure pixels and in the case of linear models, 
endmembers and abundances can be simultaneously estimated by solving a constrained or penalized non-negative matrix
factorization problem (NMF) \cite{lee_algorithms_2000}.
For example, the authors of \cite{zhuang_regularization_2019} have proposed a formulation that
involves a data fidelity term and a minimum volume
regularization term on endmembers,
whose minimization consists in alternating between solving for endmembers and abundances.

In this work, we do not assume the existence of pure pixels as they are often
missing in real data, since, for instance, the spectral signatures of endmembers in
HSI can be significantly affected by various changes in atmospheric,
illumination, and environmental conditions within the scene~\cite{borsoi_spectral_2021}. 
We mitigate the effect of spectral variability by (i)
normalizing each pixel by the $\ell_2$-norm of its spectrum as a pre-processing
step and (ii) modeling endmembers as convex combinations of pixels present
in the scene.
Not only HSI pixels are linear combinations of the estimated endmembers under 
the linear mixing model, but the estimated endmembers are also
convex
combinations of pixels. 
This corresponds to the archetypal analysis (AA) formulation introduced by Cutler and
Breiman in \cite{cutler_archetypal_1994}.
Compared to plain NMF, AA provides greater interpretability given that each
endmember can be traced back to pixels present in the HSI.
In addition, since the estimated endmembers spectral signatures generally
correspond to
averaging
the contributions of several pixels, the resulting estimation appears to be more robust
to noise and spectral variability than pure pixel methods.

\subsection{Contributions and novelties}

  \begin{enumerate}
  \item We propose a new hyperspectral unmixing algorithm relying on
    entropic gradient descent for archetypal analysis. Our approach (i) provides better
    solutions for hyperspectral unmixing than traditional alternating optimization schemes based on
    projected gradient methods or active set algorithms, and (ii) allows
    more efficient GPU implementations.
  \item The efficiency of our method enables us to make a key practical contribution, consisting of an ensembling mechanism along with an appropriate model selection
    procedure, which makes our method almost parameter-free and thus easy to use (the only sensitive parameter
    is the number of endmembers we want to estimate).
  \item Our approach, available in an open-source package\footnote{Code is available at \href{https://github.com/inria-thoth/EDAA}{https://github.com/inria-thoth/EDAA}}, outperforms state-of-the-art matrix factorization and deep learning methods on six standard real datasets.
  \end{enumerate}

The remainder of this paper is organized as follows. Section~\ref{sec:relwork}
presents related works for unsupervised linear hyperspectral unmixing.
Section~\ref{sec:method} introduces our method.
Section~\ref{sec:exp} presents experimental results highlighting the performance
of our proposed approach.
Finally, we conclude the article and underline future research directions in
Section~\ref{sec:ccl}.

\section{Related work on Unsupervised Linear Hyperspectral Unmixing}
\label{sec:relwork}

In this section, we present the framework of
non-negative matrix factorization (NMF) before introducing archetypal analysis
(AA). Finally, we mention widely used deep learning architectures that
tackle blind hyperspectral unmixing.

\paragraph{Non-negative matrix factorization (NMF)}

NMF \cite{paatero_positive_1994, lee_algorithms_2000} is popular for hyperspectral unmixing. It consists of factorizing a
matrix representing the HSI signal into the product of two matrices with
non-negative entries, where one factor carries the endmembers whereas the other
one represents the abundances for each pixel (typically with the constraint that abundances sum to one).
Several variants of NMF have been proposed for this task.
For instance, the method of minimum volume constrained non-negative matrix
factorization (MVC-NMF) of \cite{miao_endmember_2007} uses a minimum volume term for endmembers that effectively gets rid of the pure pixel
assumption. In minimum dispersion constrained NMF
(MiniDisCo)~\cite{huck_minimum_2010}, the regularization function is called \emph{dispersion}, which encourages endmembers with minimum variance to
avoid degenerate solutions and improve the unmixing robustness to flat spectra.
Other approaches design regularization functions for abundances such
as~\cite{zymnis_hyperspectral_2007, yang_blind_2010}.

\paragraph{Archetypal analysis (AA)}
AA was first introduced in~\cite{cutler_archetypal_1994} and
consists in modeling individual pixels present in the HSI as a linear
mixture of archetypes (here the endmembers).
Interestingly, archetypes are defined as convex
combinations of the individual pixels present in the HSI.
Motivated by the good interpretability of AA, Zhao \etal~\cite{zhao_multiple_2015} have proposed a kernelized variant,
which enables greater modeling flexibility at the cost of an extra
hyper-parameter (the bandwidth of the RBF kernel functions).
Notably, they use the relaxation form introduced in
\cite{morup_archetypal_2012} to handle the cases where endmembers are located outside of the convex hull of the data.

Inspired by the robust AA formulation introduced in \cite{chen_fast_2014} that
considers the Huber loss instead of the squared loss to reduce the impact of
noise and outliers, Sun \etal \cite{sun_pure_2017} have introduced a robust kernel archetypal analysis
(RKADA) method for blind hyperspectral unmixing.
Their approach refines the standard AA formulation by adding a binary
sparse constraint on the pixels contributions.
Consequently, their method ensures that each endmember actually corresponds to
real pixels rather than a sparse linear mixture of all pixels.
The resulting optimization algorithm is a block-coordinate descent scheme that
uses an active-set method \cite{nocedal_numerical_1999} to solve the smooth
least squares optimization subproblems with a simplicial constraint.
As noted by \cite{chen_fast_2014}, the active-set algorithm can be seen as an
aggressive strategy that can leverage the underlying sparsity of the subproblems solutions.

Recently, Xu \etal \cite{xu_l1_2022} have proposed an $\ell_1$ sparsity-constrained AA algorithm
to increase the sparsity of the abundances.
Rather than using an active-set algorithm, they adopt a projected gradient
method \cite{lin_projected_2007} to solve the resulting optimization subproblems inside an alternating scheme.
However their formulation leads to abundances that do not sum to one, which reduces the physical
interpretability of the unmixing results. Moreover, their approach relies on
other algorithms including PPI \cite{boardman_mapping_1995} for pre-processing
and initialization which increases the complexity of their algorithm.

\paragraph{Autoencoders (AE)}

Interest in deep learning has been rapidly growing in many fields including
remote sensing \cite{zhang_deep_2016} thanks to the increasing amount of
available data, rising computational power, and the development of
suitable algorithms.
Autoencoders for hyperspectral unmixing have been widely used since the pioneering
work of Licciardi \etal \cite{licciardi_unsupervised_2012}.
In a nutshell, the encoder transforms an input into abundances maps that are then
linearly decoded by the decoder.  The encoder activations correspond to the
fractional abundances while the parameters of the linear decoder layer
correspond to the endmembers spectra.
While the AE described above corresponds to a basic fully
connected architecture, refinements have been introduced such as using a loss functions
that involves the spectral angle distance and a dedicated sparsity penalty term
as in \cite{ozkan_endnet_2018}, or convolutions to take
advantage of the spatial structure in HSI \cite{palsson_convolutional_2020}. See \cite{palsson_blind_2022} for a comprehensive study on the various existing
autoencoder architectures for blind hyperspectral unmixing.
Finally, it is worth mentioning the recent effort by Rasti \etal
\cite{rasti_misicnet_2022} to incorporate a geometrically
motivated penalty into deep learning-based approaches. Their method, called
minimum simplex convolutional network (MiSiCNet), uses both geometrical and spatial information, by means of a
convolutional encoder-decoder architecture, to tackle unsupervised hyperspectral unmixing.

Although the surge of deep learning methods has led to improving overall
unmixing performances, training a single network per image remains costly due
to the required extensive hyperparameters search.
Moreover, training such networks requires GPU, and yet it can be considerably slower
than traditional methods that run on CPU.

\section{Methodology}

In this section, we present our model formulation before describing its
optimization.  Next, we mention implementation details required to run our
approach.  Finally, we explain how to leverage our efficient GPU implementation
and propose a procedure to make our model robust to hyper-parameter choices
and thus easy to use in practice.

\label{sec:method}

\subsection{Model formulation} \label{subsec:assumptions}

Let $\X$ in $\Real^{L \times N}$ be a hyperspectral image (HSI)
where $L$ is the number of channels and $N$ is the number of pixels.
We assume a linear mixing model such that
\begin{equation}
\label{eq:LMM}
\X = \E \A + \N,
\end{equation}
where $\E$ in
$\Real^{L \times p}$ is the \emph{mixing matrix} composed of the discretized spectra of
$p$ endmembers over $L$ channels, and $\A$ in $\Real^{p \times N}$ is the
\emph{abundance matrix} that describes, for each pixel, the fraction relative to each endmember.
Finally, $\N$ represents some classical noise occuring in hyperspectral imaging.

We are interested in tackling the \emph{blind} unmixing setting where both the
mixing and abundance matrices are unknown. However, we assume that the number $p$
of endmembers is known.
Since $\E$ represents the endmembers reflectance over~$L$ channels, it follows that its
elements should be non-negative, \ie, $\E \geq 0$. 
This is also the case for $\A$, whose columns  represent the abundances for each pixel,
which in addition, should sum to one.
In other words,  each column of $\A = \left[\abold_{1},
  \ldots, \abold_{N} \right]$ belongs to the simplex $\Delta_p$ defined as:
\begin{equation} \label{eq:simplex}
  \Delta_p \triangleq \left\{\abold \in \Real^{p} \ \text{s.t.} \ \abold \geq 0
    \  \text{and} \ 
    \sum_{j=1}^p \abold_j = 1 \right\}.
\end{equation}
The previous model and constraint yields the classical optimization problem
\begin{argmini}
  {\E,\A}{\frac{1}{2}\|\X - \E \A\|_F^2.}{\label{blind1}}{}
  \addConstraint{\E}{ \geq 0}
  \addConstraint{\abold_{i}}{\in \Delta_p \ \text{for} \ 1 \leq i \leq N},
\end{argmini}
which is a variant of non-negative matrix factorization.

The archetypal analysis formulation we consider adds a constraint 
and forces the endmembers to be convex combinations of the pixels present in
$\X$. Formally, it simply means that there exists a 
matrix $\B$ in $\Real^{N \times p}$ such that $\E = \X\B$ and the columns of $\B$
are in the simplex $\Delta_N$ (their entries are non-negative and sum to one).
This formulation yields the archetypal analysis formulation~\cite{cutler_archetypal_1994}:
\begin{argmini}
  {\A,\B}{\frac{1}{2}\|\X - \X \B \A\|_F^2,}{\label{AA}}{}
  \addConstraint{\abold_{i}}{\in \Delta_p \ \text{for} \ 1 \leq i \leq N}
  \addConstraint{\bbold_{j}}{\in \Delta_N \ \text{for} \ 1 \leq j \leq p}
\end{argmini}
where $\A = [\abold_{1}, \ldots, \abold_{N}]$ and $\B = [\bbold_{1}, \ldots,
\bbold_{p}]$.

\subsection{Optimization} \label{subsec:optim}

Minimizing (\ref{AA}) is difficult since the objective function is not jointly
convex in $(\A, \B)$.  However, it is convex with respect to (w.r.t.) one of
the variables when the other one is fixed \cite{morup_archetypal_2012}.
Consequently, it is natural to consider an alternating minimization scheme
between $\A$ and $\B$, which is guaranteed to asymptotically provide a
stationary point of the problem~\cite{bertsekas_nonlinear_1997}.  Yet, 
because of the non-convexity of the objective, the choice of optimization
algorithm is important as different stationary points may not have the same
quality in terms of statistical estimation. In other words, different
optimization procedures may have a different ``implicit bias'', a phenomenon
that is currently the focus of a lot of attention in machine
learning~\cite{pesme_implicit_2021}, especially for deep learning models, suggesting that it may also be key for
blind HU.

In this paper, we adopt an optimization scheme called entropic gradient descent
which has shown better theoretical properties in terms of convergence rates
than gradient descent when optimizing over the simplex~\cite{beck_mirror_2003}.
Our second motivation was the simplicity of the algorithm, which does not
require performing orthogonal projections on the simplex, nor dealing with
complicated active-set rules as in~\cite{chen_fast_2014}.  This enables us to
take advantage of modern GPUs.
We now present mathematical details before discussing implementation.

\subsubsection{Entropic descent algorithm (EDA)} \label{subsubsec:eda}
As noted in \cite{beck_mirror_2003}, the entropic descent algorithm is
simply a gradient descent method with a particular choice of a Bregman-like
distance \cite{bregman_relaxation_1967} generated by a specific function, here
the negative entropy.  As explained in \cite{teboulle_entropic_1992}, the
choice of an appropriate distance-like function tailored to the geometry of the
constraints, here the simplex,
provide theoretical benefits in terms of convergence rates.
The minimization over the simplex is precisely the reason why we adopt the
negative entropy to derive the updates of the alternating minimization scheme.
Formally, the negative entropy function~$h$ is defined as follows: for $x$ in $\Real^d$,
\begin{equation} \label{eq:negent}
  \quad h(x) = \sum_{i=1}^d x_i \text{ln}(x_i) \; \text{if} \; x \in \Delta_d, \quad +\infty \; \text{otherwise},
\end{equation}
with the convention that $0 \; \text{ln}\; 0 \equiv 0$.

$h$ exhibits several desirable properties, including convexity on $\Delta_d$.
This enables us to consider $D_h$, the Bregman divergence \cite{bregman_relaxation_1967} w.r.t. $h$, defined,
for $x$ and $y$ in $\Real^d$
\begin{equation} \label{eq:bregman}
\begin{aligned}
  D_h(x, y) &= h(x) - h(y) - \nabla h
  (y)^\top (x - y),
\end{aligned}
\end{equation}
which is also called the Kullback-Leibler divergence. By convexity of $h$, we naturally have $D_h(x,y) \geq 0$. 
We start by considering the following generic optimization problem:
\begin{equation} \label{eq:deff}
  \min_{\zbf \in \Delta_d} f(\zbf),
\end{equation}
where $f$ is a convex Lipschitz continuous function with a gradient at $\zbf \in
\Delta_d$ denoted by $\nabla f(\zbf)$ and $\Delta_d$ corresponds to the $d$-dimensional
simplex~(\ref{eq:simplex}).

The algorithm to solve (\ref{eq:deff}) we consider performs the following iterates,
given $\zbf^k \in \Delta_d$, 
\begin{equation} \label{eq:EMDA}
\zbf^{k+1} \leftarrow \arg \min_{\zbf \in \Delta_d} \left\{\nabla f(\zbf^k)^\top(\zbf - \zbf^k) + \frac{1}{\eta_k} D_h(\zbf, \zbf^k) \right\}.
\end{equation}
If $D_h$ was simply a squared Euclidean norm, we would recover a projected gradient
descent algorithm. By using instead the Bregman distance function $D_h$ derived
from the negative entropy, we obtain the entropic descent method.
Here $D_h$ measures the distance between two vectors in $\Delta_d$.
As such, the next iterate $\zbf^{k+1}$ should aim for the optimal balance
between taking a gradient step and moving the least from the current iterate
$\zbf^k$ according to the geometry induced by~$h$, with $\eta_k$ controlling this trade-off.

We will now see how the negative entropy $h$ yields  
explicit steps that effectively enforce the simplicial constraints.
Indeed, after a short classical calculation (see, for instance~\cite{beck_mirror_2003}),  
it is possible to show that the update~(\ref{eq:EMDA}) is equivalent to the
following one: for all $j$ in $\{1, \ldots, d\}$,
\begin{equation} \label{eq:update}
  \zbf_j^{k+1} = \frac{\zbf_j^k e^{-\eta_k \nabla f(\zbf^k)_j}}{\sum_{l=1}^d \zbf_l^k e^{-\eta_k \nabla f(\zbf^k)_l}},
\end{equation}
where $\zbf_j^k$ is the $j$-th entry of the vector $\zbf^k$ and similarly, 
$\nabla f(\zbf^k)_j$ is the $j$-th entry of $\nabla f(\zbf^k)$.

It is thus easy to see that the iterates $\zbf^{k}$ stay in the simplex~$\Delta_d$,
and it is possible to show (see~\cite{beck_mirror_2003}) that 
the sequence $(\zbf^k)_{k}$ converges to
the set of solutions of~(\ref{eq:deff}) with the appropriate step-sizes $\eta_k$.
Interestingly, the update~(\ref{eq:update}) can be implemented efficiently by using
the softmax function, assuming the entries of $\zbf^k$ are positive:
 \begin{equation} \label{eq:update_softmax}
   \zbf^{k+1} = \text{softmax}\left(\log(\zbf^k) - \eta_k \nabla f(\zbf^k)\right).
 \end{equation}
 where $\log(\zbf^k)$ is the vector carrying the logarithm of each entry of $\zbf^k$.
 This update immediately suggests a high compatibility with GPUs.

\subsubsection{Alternating optimization scheme}
We are now in shape to describe an alternating optimization scheme, consisting
of performing, alternatively, $K_1$ updates of EDA for minimizing
$\A$ when $\B$ is fixed, and vice versa using $K_2$ updates.
This strategy is presented in Algorithm \ref{alg:EDAA}.
Formally, by replacing the generic function~$f$ with the functions corresponding
to the two optimization subproblems, we obtain the following updates:
\begin{equation} \label{eq:update_A}
\A^{k+1} = \text{softmax}\left( \log(\A^k) + \eta_k \B^\top \X^\top (\X - \X \B \A^k) \right),
\end{equation}

\begin{equation} \label{eq:update_B}
  \B^{k+1} = \text{softmax}\left( \log(\B^k) + \eta_k \X^\top (\X - \X \B^k \A) \A^\top \right),
\end{equation}

where $\log(\A^k)$ is the matrix carrying the logarithm of each entry of $\A^k$
while $\text{softmax}$ is applied in parallel on the columns of $\log(\A^k) +
\eta_k \B^\top \X^\top (\X - \X \B \A^k)$. Note that when $\A$ and $\B$ are
initialized with positive values, these iterates keep them positive.

\begin{algorithm}[h]
  \caption{Entropic Descent Archetypal Analysis (EDAA)}\label{alg:EDAA}
  \begin{algorithmic}[1]
    \State \textbf{Input:} $\ell_2$-normalized data $\X$ in $\Real^{L \times N}$; $p$ (number of
    endmembers); $T$ (number of outer iterations); $K_1$ (number of inner
    iterations for $\hat{\A}$); $K_2$ (number of inner iterations for $\hat{\B}$).
    \State Initialize $\hat{\A}  \in \Real^{p \times N}$ using (\ref{eq:initA0}).
    \State Initialize $\hat{\B} \in \Real^{N \times p}$ using (\ref{eq:initB0}).

    \State Set $\eta_1$ according to (\ref{eq:eta0A}).
    \State Set $\eta_2$ according to (\ref{eq:eta0B}).
    \For{$t = 1,\ldots, T$}
      \For{$k = 1,\ldots,K_1$} 
      \State $\hat{\A} \gets \text{softmax}\left( \log(\hat{\A}) + \eta_1
        \hat{\B}^\top \X^\top (\X - \X \hat{\B} \hat{\A}) \right)$\\
      \Comment{$\log$ is applied element-wise;}\\
      \Comment{$\text{softmax}$ is applied along the first dimension.}
      \EndFor
      \For{$k = 1, \ldots, K_2$}
      \State $\hat{\B} \gets \text{softmax} \left( \log(\hat{\B}) + \eta_2
        \X^\top (\X - \X \hat{\B} \hat{\A}) \hat{\A}^\top \right)$
      \EndFor
    \EndFor
    \State $\hat{\E} \gets \X \hat{\B}$
    \State \textbf{Return} $\hat{\E}$, $\hat{\A}$ \Comment{Estimated endmembers, abundances.}
  \end{algorithmic}
\end{algorithm}

\subsection{Implementation details}

\paragraph{Normalization}

The input image $\X = [\xbf_1, \ldots, \xbf_N]$ is $\ell_2$-normalized for
each pixel: for all $i$ in $\{1, \ldots, N\}$,

\begin{equation} \label{eq:normalization}
  \xbf_i \gets \frac{\xbf_i}{||\xbf_i||_2},
\end{equation}

where $\xbf_i$ denotes the $i$-th pixel. This step is important to gain
invariance to illumination changes.

\paragraph{Initialization}

We initialize the abundance matrix $\A$ uniformly,

\begin{equation}
  \label{eq:initA0}
  \A^0 = \frac{1}{p} \1_p \1_N^T,
\end{equation}
where $\1_d$ denotes a $d$-dimensional vector of ones.
This corresponds to the maximal entropy configuration for each pixel.
The entropy for each pixel will naturally decrease as a result of the
optimization, but the high entropy of the initialization will have a regularization
effect.

The initialization of the pixel contribution matrix $\B$ is then also close to 
uniform. Nevertheless we, introduce a small random perturbation which is necessary
to break the symmetry between the columns of $\B$ (otherwise, the updates of $\A$ and~$\B$ will keep
them invariant).
Concretely, the entries of $\B$ are randomly sampled according to the uniform distribution
on $[0,1]$, $\mathcal{U}_{[0,1]}$.
Next, they are rescaled by a factor $0.1$.
Finally, we apply the softmax function on each column so that the columns of $\B$ belong
to the simplex $\Delta_N$, for $j$ in $\{1, \ldots, p\}$,

\begin{equation}
  \label{eq:initB0}
  \bbold^0_j = \text{softmax}(0.1 \; \ubold),
\end{equation}
where $\ubold \sim \mathcal{U}_{[0,1]^N}$. In practice, we observe that such an initialization leads to 
a matrix $\B^0$ that is very close to an uniform initialization
  $\frac{1}{N} \1_N \1_p^T$.

\paragraph{Step sizes}
We use constant step sizes $\eta_1$ and $\eta_2$, for $\A$ and $\B$
respectively.

\begin{equation}
  \label{eq:eta0A}
  \eta_1 = \frac{\gamma}{\sigma^2_{\max}},
\end{equation}
where $\gamma$ is a value in $S = \{0.125, 0.25, 0.5, 1, 2, 4, 8\}$ and $\sigma_{\max}$ is the largest singular value of the matrix $\X \B^0$.
We recover the classical convergence of gradient descent with fixed step size \cite{nesterov_introductory_2003} up
to the factor $\gamma$, since $\sigma^2_{\max}$ corresponds to the Lipschitz constant of the sub-problem
related to (\ref{AA}) when minimizing w.r.t $\A$, $\B$ being fixed.
Having $\gamma$ in $S$ allows us to use slightly different step sizes and yields
better performance in practice. Note that our model selection procedure, presented later, will automatically choose
the right parameter $\gamma$, thus removing the burden for the user of having to deal with an extra hyper-parameter.
Finally, $\eta_2$ is simply a rescaled version of $\eta_1$ to account for the matrices
being of transposed dimensions:

\begin{equation}
  \label{eq:eta0B}
  \eta_2 = \sqrt{\frac{p}{N}} \; \eta_1.
\end{equation}

\paragraph{Hyperparameters}

For all experiments, we set $T=100$ and $K_1 = K_2 = 5$ as it provides a good trade-off between
convergence speed and unmixing accuracy. 
Note also that these hyper-parameters are robust to different real datasets as detailed in section \ref{sec:exp}.

\subsection{Model selection procedure}

As stated above, the initialization of the matrix $\B$ is random, leading to
different solutions for each run of the algorithm since the overall
optimization problem is non convex. Besides, we allow for different step-sizes
$\gamma$, which we draw randomly from the set $S$ in practice,
Since the convergence of the algorithm is very fast (see
experimental section for concrete benchmarks), we are able to provide a large diversity of solutions
given a dataset by running $M$ times our method with different random seeds, while keeping the global computational complexity reasonable.
A major question we address next is then \emph{how to select optimally the best solution
in terms of unmixing accuracy}.

\begin{figure*}[h]
  \centering
  \subfloat[Samson]{\includegraphics[width=0.33\textwidth]{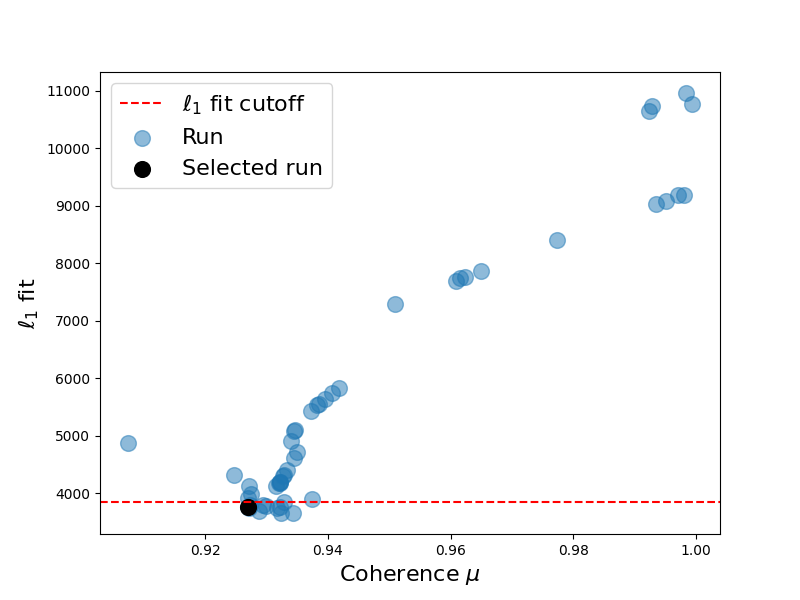}}
  \hfil
  \subfloat[Jasper Ridge]{\includegraphics[width=0.33\textwidth]{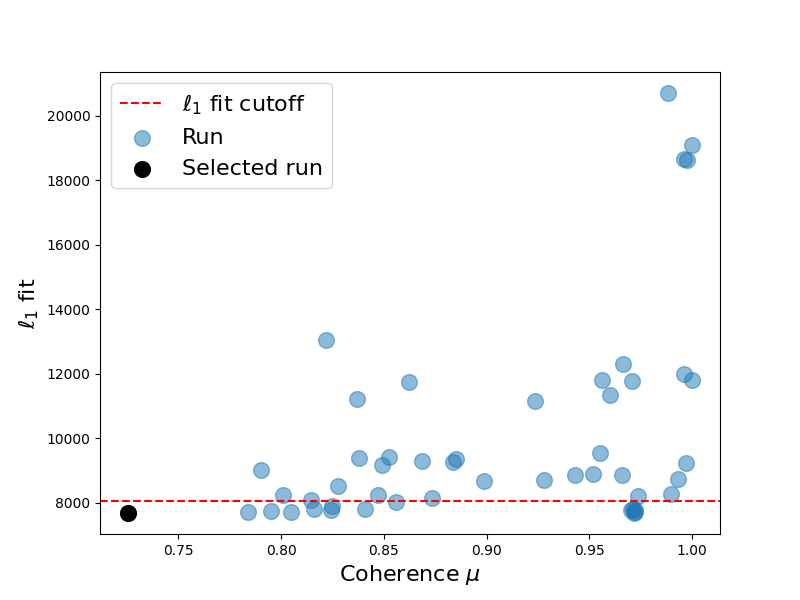}}
  \hfil
  \subfloat[APEX]{\includegraphics[width=0.33\textwidth]{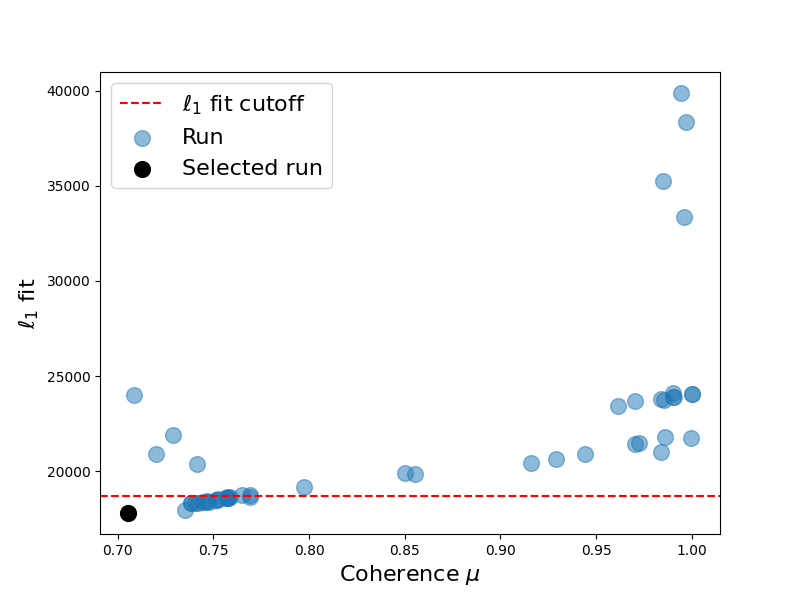}}
  \caption{Illustration of the model selection procedure on three datasets
    using $M=50$ runs. Runs are illustrated by blue dots and the
    selected one is in black. The selected run is the one with lowest coherence
    $\mu$ under the dashed red line representing the $\ell_1$ fit threshold, see Alg.~\ref{alg:criterion}.}
  \label{fig:MSP}
\end{figure*}

For this, we take inspiration from classical model selection
and sparse estimation theory~\cite{hastie_elements_2009}. First, we measure the fit of each solution in terms
of residual error $\|\X - \X\A\B\|_1$, choosing the $\ell_1$-norm which is
known to be more robust to outliers than the mean squared error. Second, we
\emph{select} the set of solutions that are in the same ball park as the best
solution we have obtained in terms of $\ell_1$ fit.
This selection process is illustrated by the red dotted line in Figure~\ref{fig:MSP},
while the precise criterion is described in Algorithm~\ref{alg:criterion}.

From the subset of solutions with good fit, we then choose the one whose endmembers have 
the best incoherence, a desired property, which is classical in the theory of
sparse estimation~\cite{elad_generalized_2002,
gribonval_sparse_2003,mairal2014sparse}. Indeed, dictionaries (here endmembers) with more incoherence
will benefit from better theoretical guarantees in terms of estimation of abundances, making it
a natural criterion for model selection in the context of unmixing.
Formally, the coherence is simply defined as the maximal pairwise spectral
correlation between the estimated endmembers. More precisely, for $\hat{\E} = [\hat{\ebf}_1, \ldots, \hat{\ebf}_p]$
the endmembers matrix, the coherence $\mu$ is defined as:
\begin{equation} \label{eq:coherence}
\mu = \max_{k \neq k'}\langle \hat{\ebf}_k, \hat{\ebf}_{k'} \rangle.
\end{equation}
To the best of our knowledge, this is the first time the coherence $\mu$ is used
as a model selection criterion for archetypal analysis. Our experiments, see
next section, show that it is highly effective.

In summary, we automatically select the model whose endmembers have the lowest maximal pairwise
spectral correlation among the ones that have a good $\ell_1$ fit.
This strategy is illustrated in figure \ref{fig:MSP} and described in Algorithm \ref{alg:criterion}.
In the experiments, the number of runs $M$ was set to 50.

\begin{algorithm}[h]
  \caption{Model Selection Procedure}\label{alg:criterion}
  \begin{algorithmic}[1]
    \State \textbf{Input:} $M$ (number of runs); $\ell_2$-normalized data $\X$ in $\Real^{L \times N}$; $p$ (number of
    endmembers); $T$ (number of outer iterations); $K_1$ (number of inner
    iterations for $\hat{\A}$); $K_2$ (number of inner iterations for $\hat{\B}$).
    \For{$m = 1, \ldots, M$} 
       \State Set random seed $s_m$.
       \State $\hat{\E}_m, \hat{\A}_m \gets \text{EDAA}(\X, p, T, K_1, K_2, s_m)$ 
       \Comment{See (\ref{alg:EDAA})}
       \State $\text{fit}_m \gets ||\X - \hat{\E}_m \hat{\A}_m||_1$
       \State Compute coherence $\mu_m$ on $\hat{\E}_m$.
       \Comment{See (\ref{eq:coherence})}
    \EndFor
    \State $\text{fit}_{\min} \gets \min \{ \text{fit}_1, \ldots, \text{fit}_M \}$
    \State $\mathcal{I} \gets \{m \; | \; \text{fit}_m \leq 1.05 \times \text{fit}_{\min} \}$
    \Comment{Subset of models.}
    \State $\text{best} \gets \arg \min \{\mu_i, i \in \mathcal{I}\}$
    \State \textbf{Return:} $\hat{\E}_{\text{best}}$, $\hat{\A}_{\text{best}}$.
  \end{algorithmic}
\end{algorithm}

\section{Experiments}

\label{sec:exp}

We have performed experiments on six standard real datasets whose descriptions are given below.

\subsection{Hyperspectral data description}

\begin{enumerate}

\item Samson: 
  The Samson\footnote{downloaded at \href{https://rslab.ut.ac.ir/data}{https://rslab.ut.ac.ir/data}} hyperspectral
image is a 95x95 pixels sub-region of a larger image
captured using 156 bands spanning from 401 to 889 nm.
Three main materials have been identified: Tree, Soil and Water. 
Note that we use a different ground truth from \cite{rasti_misicnet_2022} that 
we selected for its sharper details on the abundances.

\item Jasper Ridge:
\addtocounter{footnote}{-1}
The Jasper Ridge\footnotemark hyperspectral
image is a 100x100 pixels sub-region of a larger image initially captured
using 224 bands spanning from 380 to 2500 nm.
In total, 198 bands remain as 26 were removed as a pre-processing step due to
dense water vapor and atmospheric effects.
Four main materials have been identified: Tree, Dirt, Water and Road.

\item Urban4 and Urban6:
\addtocounter{footnote}{-1}
The Urban\footnotemark hyperspectral image is a 307x307 pixels image collected by the
Hyperspectral Digital Image Collection Experiment (HYDICE) \cite{rickard_hydice_1993} sensor using
210 bands spanning from 400 to 2500 nm.
In total, 162 bands remain as 48 were removed as a pre-processing step due to
dense water vapor and atmospheric effects.
There exists three versions of this dataset w.r.t. the number of endmembers.
In this study, we focus on the two extremes: Urban4 contains 4 endmembers
(Asphalt Road, Grass, Tree and Roof) and
Urban6 contains two additional materials: Dirt and Metal, making it more challenging.

\item APEX:
The APEX \cite{schaepman_advanced_2015} hyperspectral image that we consider in
this paper is a 111x122 pixels cropped region\footnote{downloaded at \href{https://github.com/BehnoodRasti/MiSiCNet}{https://github.com/BehnoodRasti/MiSiCNet}} of a larger image captured over
285 bands spanning from 413 to 2420nm.
Four main materials were identified: Road, Tree, Roof and Water.

\item Washington DC Mall:
\addtocounter{footnote}{-1}
The Washington DC Mall (WDC) hyperspectral dataset\footnotemark consists in a
319x292 pixels image captured by the HYDICE \cite{rickard_hydice_1993} sensor
over 191 bands spanning from 400 to 2400 nm.
Six main materials were identified: Grass, Tree, Roof, Road, Water and Trail.
\end{enumerate}

According to \cite{zhu_hyperspectral_2017} (Samson, Jasper Ridge and Urban) and
\cite{rasti_misicnet_2022} (APEX and WDC), the endmembers
spectra were manually selected from the images and the ground truth abundances
were set by FCLSU.
Illustrations of the datasets and their ground truth endmembers are available in
the supplementary material.

\subsection{Experimental setup}

We compare our approach to five competitive methods from different unmixing
categories:

\begin{itemize}
  \item Geometrical unmixing baseline: FCLSU \cite{heinz_fully_2001} using VCA
    \cite{nascimento_vertex_2005} to extract endmembers.
    Our implementation of the FCLSU algorithm uses the \emph{DecompSimplex}
    routine implemented in SPAMS\footnote{\href{http://thoth.inrialpes.fr/people/mairal/spams/}{http://thoth.inrialpes.fr/people/mairal/spams/}}.
    This method relies on the active-set algorithm \cite{nocedal_numerical_1999}
    that enables significantly faster convergence than generic quadratic
    programming solvers by leveraging the underlying sparsity of the abundances
    as noted by \cite{chen_fast_2014}.

  \item Deep learning unmixing: Endnet\footnote{implementation at \href{https://github.com/burknipalsson/hu\_autoencoders}{https://github.com/burknipalsson/hu\_autoencoders}}
    \cite{ozkan_endnet_2018} using VCA \cite{nascimento_vertex_2005} to
    initialize the endmembers and MiSiCNet\footnote{implementation at
      \href{https://github.com/BehnoodRasti/MiSiCNet}{https://github.com/BehnoodRasti/MiSiCNet}}
    \cite{rasti_misicnet_2022}.

  \item NMF-based blind unmixing: non-negative matrix factorization quadratic
    minimum volume (NMF-QMV)\footnote{implementation at
      \href{https://github.com/LinaZhuang/NMF-QMV\_demo}{https://github.com/LinaZhuang/NMF-QMV\_demo}}
    \cite{zhuang_regularization_2019} using the \emph{boundary} term as the
    quadratic minimum volume penalty and AA \cite{cutler_archetypal_1994} using
    the implementation from \cite{chen_fast_2014} developed in SPAMS.
\end{itemize}

To quantitatively evaluate the performance of the selected methods, we consider two metrics that are computed both globally and individually for each endmember.
On one hand, we measure the quality of the generated abundances by means of the
abundances root mean square error (RMSE) in percent between the ground truth and the
estimated abundances:

\begin{equation} \label{eq:RMSE}
  \text{RMSE}(\A, \hat{\A}) = 100 \times \sqrt{\frac{1}{p N} \sum_{i=1}^p \sum_{j=1}^N \left( \A_{i, j} - \hat{\A}_{i, j} \right)^2}.
\end{equation}

On the other hand, we assess the quality of the estimated endmembers spectra by
using the spectral angle distance (SAD) in degrees between the ground truth and
the generated endmembers:

\begin{equation}  \label{eq:SAD}
  \text{SAD}(\E, \hat{\E}) = \frac{180}{\pi} \times \frac{1}{p} \sum_{i=1}^p \arccos \left( \frac{\langle \ebf_i, \hat{\ebf}_i \rangle}{||\ebf_i||_2 ||\hat{\ebf}_i||_2} \right),
\end{equation}

where $\langle . \rangle$ denotes the inner product and $\ebf_i$ denotes the
$i$-th column of $\E$, \ie the spectrum of the $i$-th endmember.

\subsection{Unmixing experiments}

\begin{table*}[h]
  \centering
  \captionof{table}{Abundances RMSE on six real datasets. The best results are
    shown in bold. The second best results are underlined.}
    \begin{tabular}{c | c | c c c c | c c}
  \toprule
  & & FCLSU \cite{nascimento_vertex_2005, heinz_fully_2001} & Endnet \cite{ozkan_endnet_2018} & MiSiCNet \cite{rasti_misicnet_2022} & NMF-QMV \cite{zhuang_regularization_2019} & AA & EDAA \\
  \hline
  \multirow{4}{*}{Samson}
  & Soil & 11.28 & 11.61 & 6.47 & 13.67 & \underline{6.16} & \textbf{5.74} \\
  & Tree & 9.13 & 7.72 & 5.38 & 8.4 & \underline{4.00} & \textbf{3.77} \\
  & Water & 5.05 & 6.86 & 3.47 & 11.61 & \textbf{2.30} & \underline{2.59}\\
  \cline{2-8}
  & Overall & 8.88 & 8.97 & 5.25 & 11.44 & \underline{4.44} & \textbf{4.24}\\
  \hline
  \hline
  \multirow{5}{*}{Jasper Ridge}
  & Dirt & 21.23 & 18.26 & 21.68 & 19.97 & \underline{10.24} & \textbf{7.32} \\
  & Road & 24.72 & 29.40 & 24.94 & 26.13 & \underline{9.79} & \textbf{7.61} \\
  & Tree & 11.20 & \underline{4.00} & \textbf{3.41} & 14.55 & 6.32 & 6.63 \\
  & Water & 13.61 & 22.38 & 7.07 & 19.81 & \underline{6.77} & \textbf{5.69} \\
  \cline{2-8}
  & Overall & 18.52 & 20.70 & 16.98 & 20.53 & \underline{8.46} & \textbf{6.85} \\
  \hline
  \hline
  \multirow{5}{*}{Urban4}
  & Road & 30.54 & 12.04 & \underline{10.30} & 20.25 & 11.12 & \textbf{8.62} \\
  & Grass & 32.99 & 17.94 & 12.35 & 20.22 & \underline{11.55} & \textbf{9.28} \\
  & Roof & 15.40 & 11.28 & 8.01 & 13.29 & \underline{7.28} & \textbf{6.37} \\
  & Tree & 20.02 & 11.69 & 8.78 & 21.56 & \underline{8.62} & \textbf{6.27} \\
  \cline{2-8}
  & Overall & 25.78 & 13.52 & 10.00 & 19.11 & \underline{9.80} & \textbf{7.75}\\
  \hline
  \hline
  \multirow{7}{*}{Urban6}
  & Road & 31.61 & \underline{17.44} & 19.18 & 24.81 & 19.23 & \textbf{11.39} \\
  & Grass & 23.62 & 31.47 & 18.84 & 27.97 & \textbf{9.73} & \underline{18.61} \\
  & Roof & 13.00 & 9.35 & \underline{7.41} & 16.39 & 10.92 & \textbf{6.01} \\
  & Tree & 16.14 & 15.50 & \underline{11.72} & 19.95 & 14.45 &\textbf{9.85} \\
  & Dirt & 25.06 & 30.67 & \underline{23.95} & 24.31 & 27.76 & \textbf{15.94} \\
  & Metal & \textbf{12.99} & 25.21 & 33.67 & \underline{13.40} & 17.53 & 15.86 \\
  \cline{2-8}
  & Overall & 21.54 & 23.09 & \underline{16.27} & 21.74 & 17.66 & \textbf{13.63} \\
  \hline
  \hline
  \multirow{5}{*}{APEX}
  & Road & 33.31 & \underline{29.32} & 32.66 & 31.43 & 34.12 & \textbf{16.54} \\
  & Tree & 20.97 & \underline{18.18} & 19.97 & 24.62 & 22.49 & \textbf{14.48} \\
  & Roof & \underline{14.15} & 15.88 & 18.42 & 14.73 & 18.54 & \textbf{11.27} \\
  & Water & 18.03 & 17.47 & \underline{16.88} & 17.99 & 16.90 & \textbf{16.83} \\
  \cline{2-8}
  & Overall & 22.77 & \underline{20.90} & 22.86 & 23.10 & 23.98 & \textbf{14.94} \\
  \hline
  \hline
  \multirow{7}{*}{WDC}
  & Grass & \underline{30.90} & \textbf{27.35} & 31.61 & 34.69 & 31.92 & 32.53 \\
  & Tree & 22.42 & 35.98 & 23.64 & \underline{19.90} & 20.13 & \textbf{11.46} \\
  & Road & 27.90 & 38.49 & 34.91 & \underline{22.49} & 38.87 & \textbf{13.97} \\
  & Roof & \textbf{8.71} & 27.04 & \underline{11.98} & 19.81 & 20.89 & 29.31 \\
  & Water & 17.76 & \underline{12.94} & 14.72 & 20.34 & 14.06 & \textbf{9.63} \\
  & Trail & 12.80 & 12.63 & \textbf{12.07} & \underline{12.36} & 15.24 & 13.19 \\
  \cline{2-8}
  & Overall & \underline{21.57} & 27.64 & 23.39 & 22.60 & 25.17 & \textbf{20.46}\\
  \bottomrule
\end{tabular}

  \label{table:RMSE}
\end{table*}

\begin{table*}[h]
  \centering
  \captionof{table}{Endmembers SAD on six real datasets. The best results are
    shown in bold. The second best results are underlined.}
    \begin{tabular}{c | c | c c c c | c c} 
  \toprule
  & & FCLSU \cite{nascimento_vertex_2005, heinz_fully_2001} & Endnet \cite{ozkan_endnet_2018} & MiSiCNet \cite{rasti_misicnet_2022} & NMF-QMV \cite{zhuang_regularization_2019} & AA & EDAA \\
  \hline
  \multirow{4}{*}{Samson}
  & Soil & 2.76 & \textbf{0.61} & 1.21 & 4.90 & \underline{0.78} & 1.64\\
  & Tree & 3.05 &  \underline{1.93} & 3.38 & 5.34 & \textbf{1.80} & 1.98\\
  & Water & 7.15 & 1.48 & 5.36 & 11.14 & \underline{1.38} & \textbf{1.31}\\
  \cline{2-8}
  & Overall & 4.32 & \underline{1.34} & 3.32 & 7.13 & \textbf{1.32} & 1.64\\
  \hline
  \hline
  \multirow{5}{*}{Jasper Ridge}
  & Dirt & 13.03 & \textbf{1.63} & 4.26 & 12.40 & \underline{2.44} & 2.74 \\
  & Road & 40.39 & 32.85 & 20.04 & 45.66 & \underline{8.00} & \textbf{3.10} \\
  & Tree & 11.16 & \underline{1.39} & \textbf{1.27} & 14.46 & 4.34 & 4.23 \\
  & Water & 13.24 & \underline{3.21} & 4.18 & 14.53 & 3.29 & \textbf{2.80} \\
  \cline{2-8}
  & Overall & 19.46 & 9.77 & 7.44 & 21.77 & \underline{4.52} & \textbf{3.22} \\
  \hline
  \hline
  \multirow{5}{*}{Urban4}
  & Road & 15.40 & 6.40 & \underline{5.73} & 14.51 & \textbf{3.73} & 6.01 \\
  & Grass & 24.18 & 3.09 & 5.84 & 16.39 & \textbf{1.81} & \underline{2.14} \\
  & Roof & 47.56 & \textbf{3.76} & 16.10 & 36.31 & 15.59 & \underline{10.49} \\
  & Tree & 19.82 & \textbf{2.32} & 4.60 & 22.48 & 3.51 & \underline{2.81} \\
  \cline{2-8}
  & Overall & 26.74 & \textbf{3.89} & 8.07 & 22.42 & 6.16 & \underline{5.36} \\
  \hline
  \hline
  \multirow{7}{*}{Urban6}
  & Road & 13.43 & \textbf{3.26} & 7.47 & 26.60 & 6.74 & \underline{4.85} \\
  & Grass & 22.30 & \underline{4.13} & 10.97 & 21.63 & 6.82 & \textbf{2.17} \\
  & Roof & 15.65 & 17.76 & \underline{13.97} & 15.55 & 18.07 & \textbf{13.70} \\
  & Tree & 20.70 & \underline{7.72} & 9.99 & 23.31 & \textbf{7.18} & 8.92 \\
  & Dirt & 69.81 & 17.42 & 19.57 & 23.60 & \textbf{11.08} & \underline{13.33} \\
  & Metal & 39.35 & \underline{7.04} & 9.75 & 68.73 & 40.97 & \textbf{4.52}\\
  \cline{2-8}
  & Overall & 30.21 & \underline{9.56} & 11.95 & 29.90 & 15.14 & \textbf{8.64} \\
  \hline
  \hline
  \multirow{5}{*}{APEX}
  & Road & 40.23 & \underline{14.46} & 33.02 & 54.53 & 37.32 & \textbf{6.83} \\
  & Tree & 14.13 & 7.53 & \underline{3.16} & 16.06 & \textbf{2.39} & 7.68 \\
  & Roof & 8.25 & \textbf{4.36} & 11.31 & 7.98 & 10.20 & \underline{7.50} \\
  & Water & 7.15 & \underline{2.83} & 6.02 & 9.71 & 2.95 & \textbf{2.21} \\
  \cline{2-8}
  & Overall & 17.44 & \underline{7.30} & 13.38 & 22.07 & 13.21 & \textbf{6.06} \\
  \hline
  \hline
  \multirow{7}{*}{WDC}
  & Grass & 17.40 & \textbf{3.54} & 17.46 & 34.36 & \underline{4.59} & 8.11\\
  & Tree & 23.73 & 12.85 & 12.36 & 17.70 & \underline{10.93} & \textbf{1.81} \\
  & Road & 32.56 & 26.76 & 33.20 & \underline{17.28} & 46.73 & \textbf{7.67}\\
  % & Grass & 17.40 & 12.43 & 16.36 & 34.36 & 4.59 & \textbf{3.64} & 4.90\\
  % & Tree & 23.73 & 8.51 & 8.66 & 17.70 & 10.93 & 6.63 & \textbf{2.90} \\
  % & Road & 32.56 & 12.28 & \textbf{5.03} & 17.28 & 46.73 & 46.2 & 6.41\\
  & Roof & 34.84 & \textbf{13.70} & \underline{28.87} & 44.87 & 43.26 & 50.97\\
  & Water & 4.78 & 1.75 & 1.57 & 19.74 & \underline{1.23} & \textbf{1.08}\\
  & Trail & 9.94 & \textbf{1.49} & \underline{3.24} & 9.60 & 5.32 & 4.75\\
  \cline{2-8}
  & Overall & 20.54 & \textbf{10.02} & 16.11 & 23.93 & 18.68 & \underline{12.40}\\
  \bottomrule
\end{tabular}

  \label{table:SAD}
\end{table*}

\begin{figure*}[h]
  \centering
  \includegraphics[width=\textwidth]{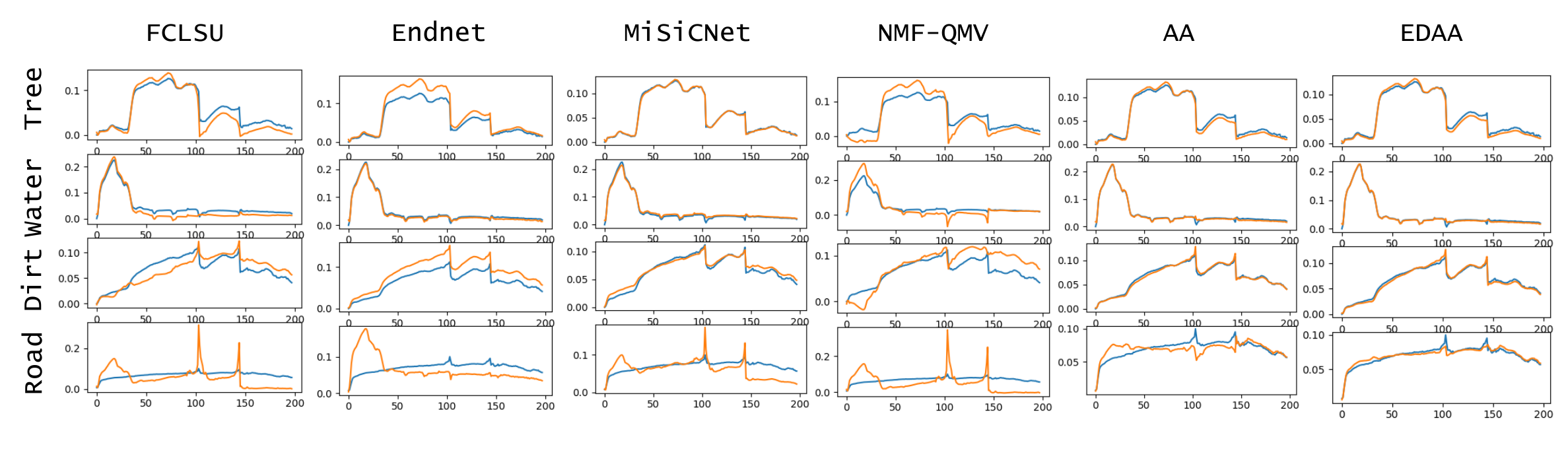}
  \caption{Estimated endmembers on the Jasper Ridge dataset.
    Ground truth endmembers are displayed in blue while their estimates are in orange.
  }
  \label{fig:JasperRidgeE}
\end{figure*}

\begin{figure*}[h]
  \centering
  \includegraphics[width=\textwidth]{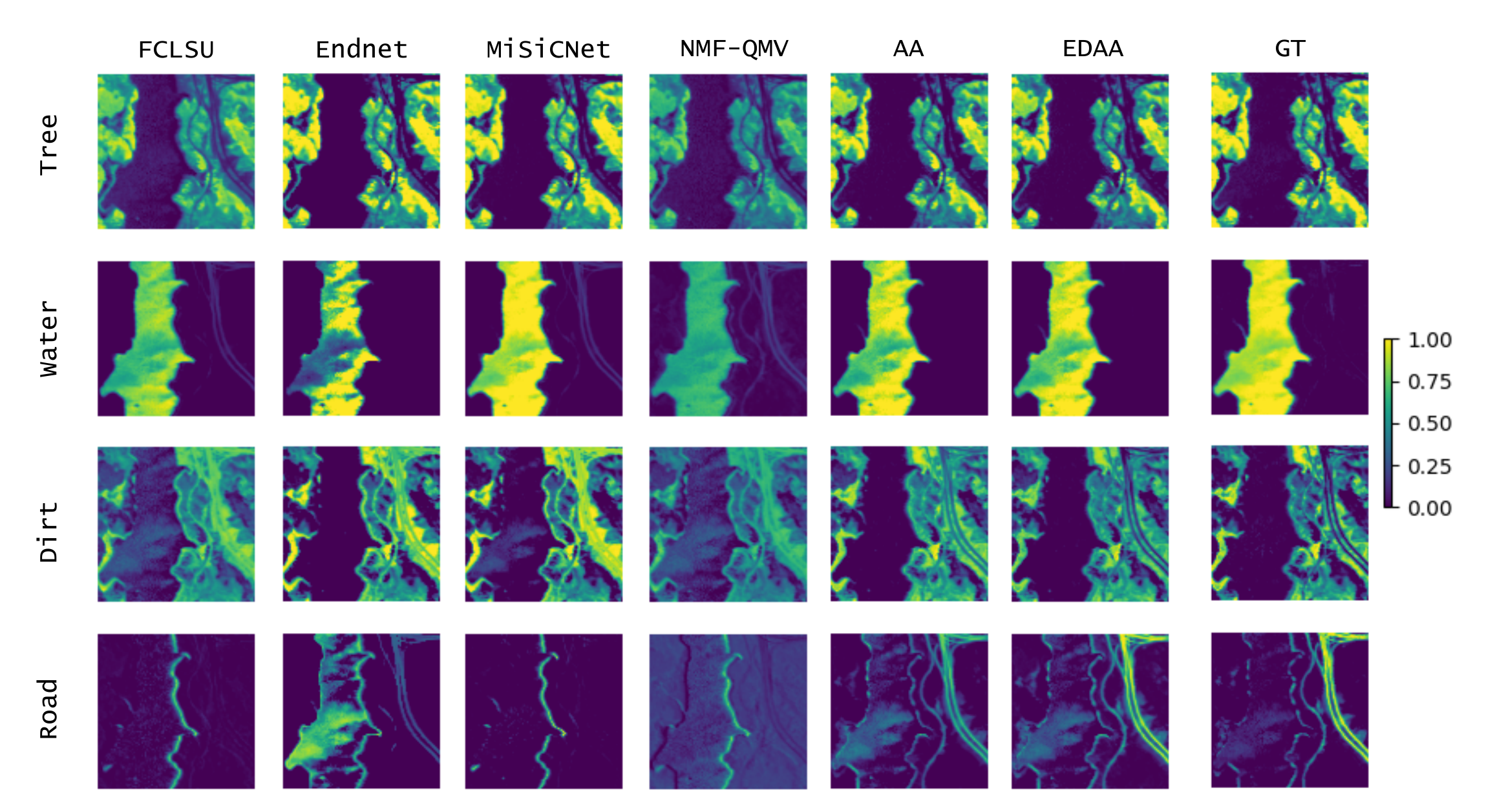}
  \caption{Estimated abundances on the Jasper Ridge dataset. The ground truth is
    displayed on the right side.}
  \label{fig:JasperRidgeA}
\end{figure*}

\begin{figure*}[h]
  \centering
  \includegraphics[width=\textwidth]{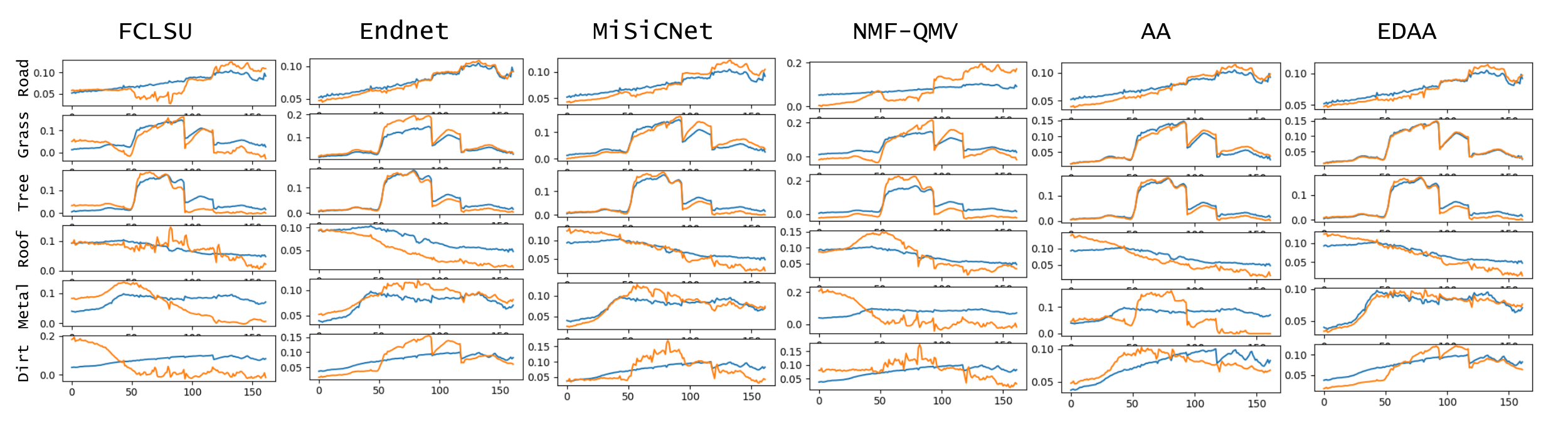}
  \caption{Estimated endmembers on the Urban6 dataset.
    Ground truth endmembers are displayed in blue while their estimates are in orange.}
  \label{fig:Urban6E}
\end{figure*}

\begin{figure*}[h]
  \centering
  \includegraphics[width=\textwidth]{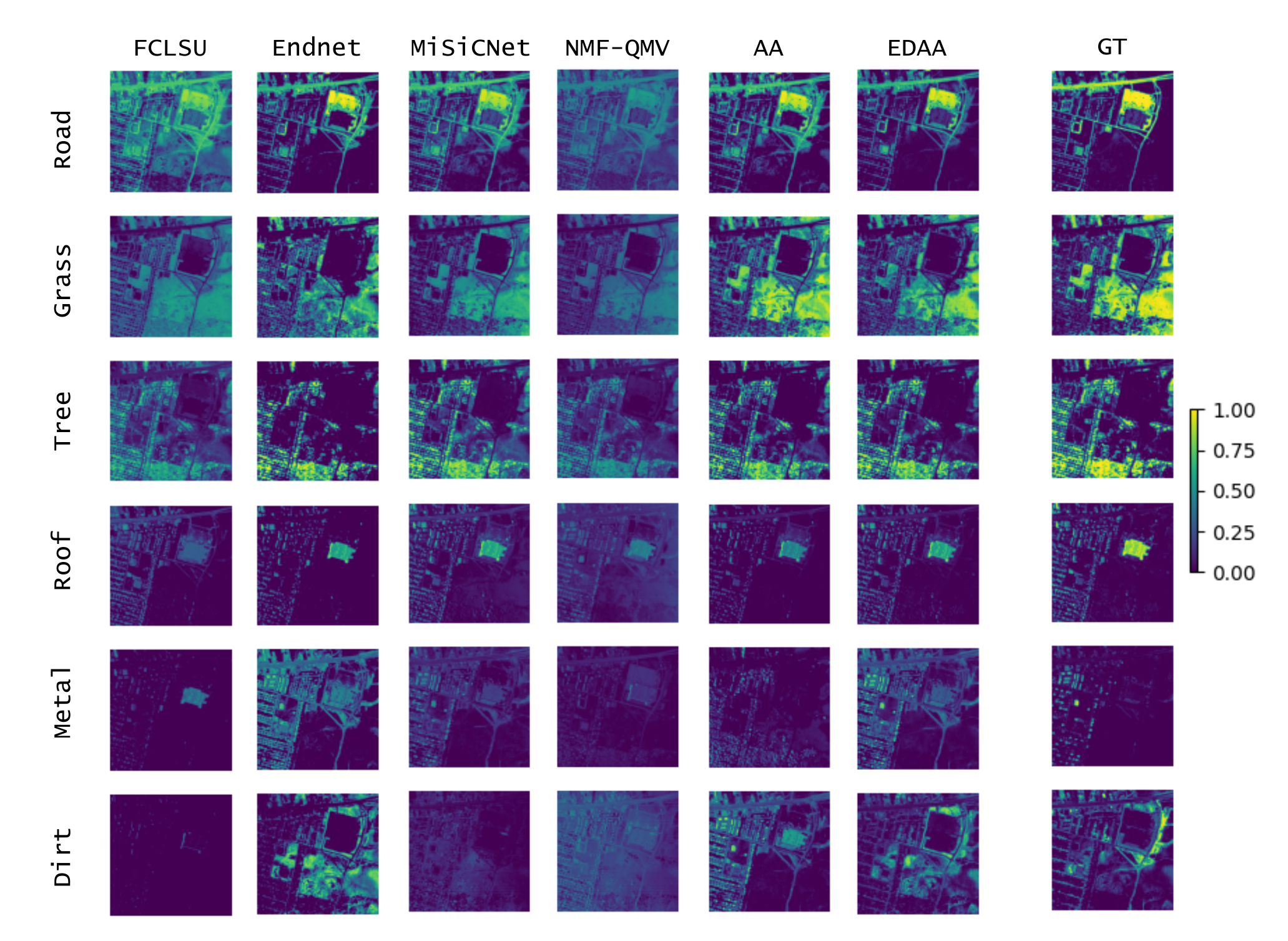}
  \caption{Estimated abundances on the Urban6 dataset. The ground truth is
    displayed on the right side.}
  \label{fig:Urban6A}
\end{figure*}

\begin{figure*}[h]
  \centering
  \includegraphics[width=\textwidth]{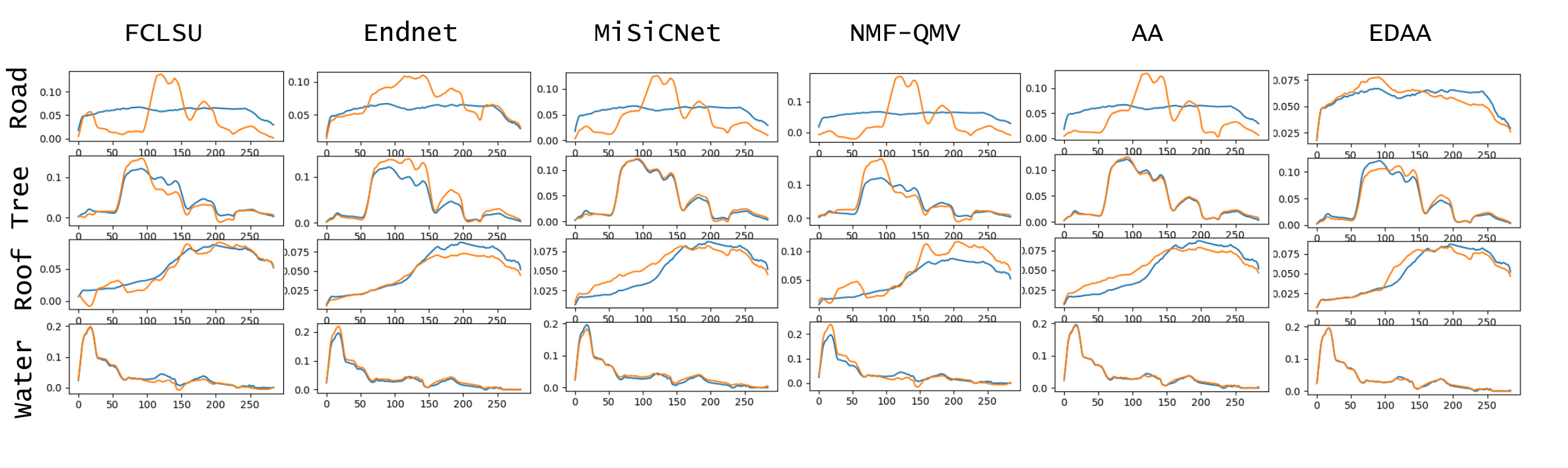}
  \caption{Estimated endmembers on the APEX dataset.
    Ground truth endmembers are displayed in blue while their estimates are in orange.}
  \label{fig:APEXE}
\end{figure*}

\begin{figure*}[h]
  \centering
  \includegraphics[width=\textwidth]{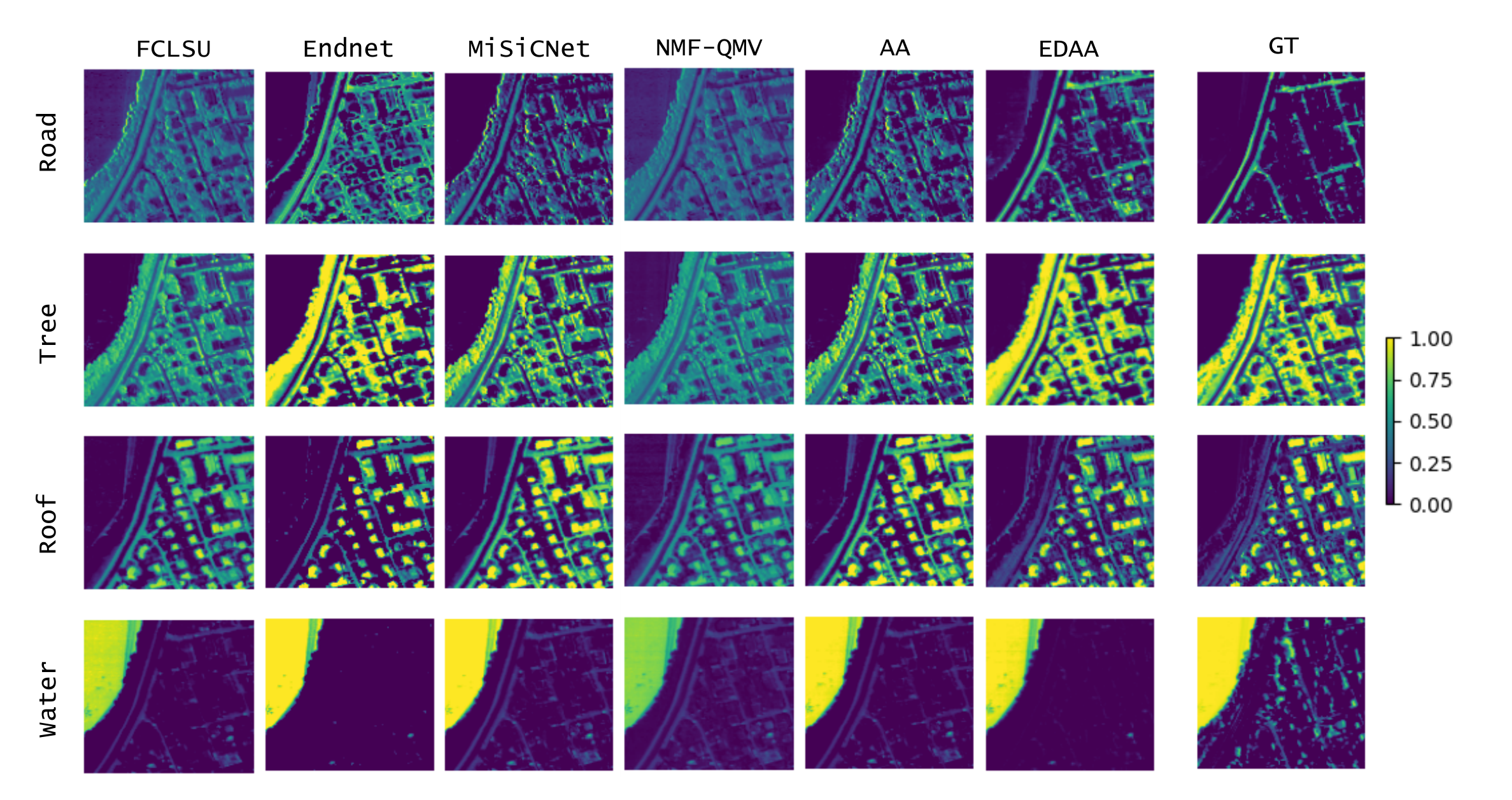}
  \caption{Estimated abundances on the APEX dataset. The ground truth is
    displayed on the right side.}
  \label{fig:APEXA}
\end{figure*}

\begin{figure*}[h]
  \centering
  \includegraphics[width=\textwidth]{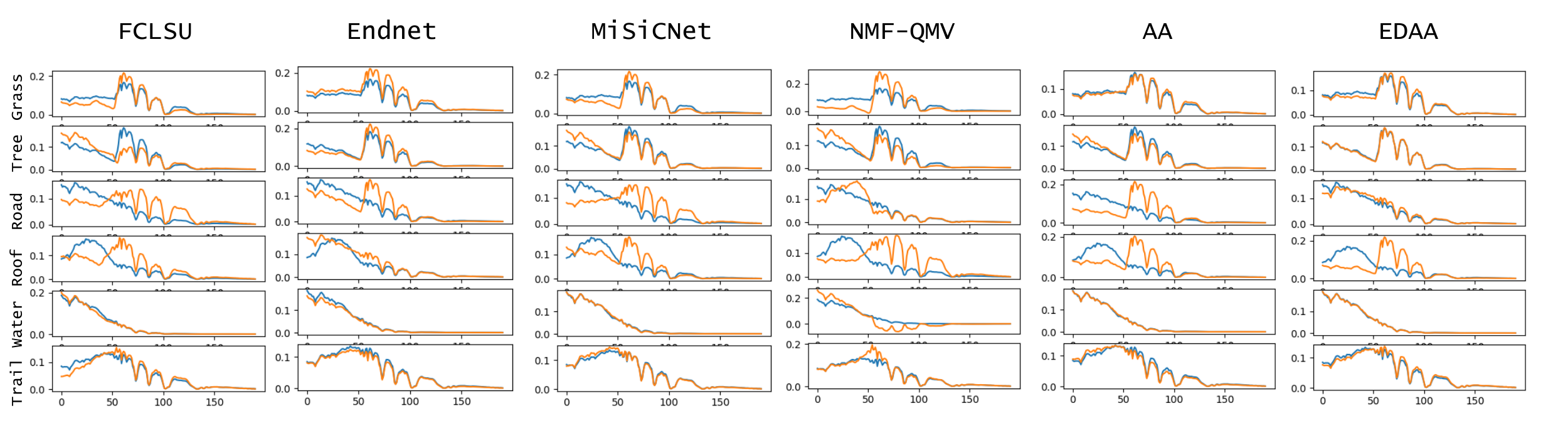}
  \caption{Estimated endmembers on the Washington DC dataset.
    Ground truth endmembers are displayed in blue while their estimates are in orange.}
  \label{fig:WDCE}
\end{figure*}

\begin{figure*}[h]
  \centering
  \includegraphics[width=\textwidth]{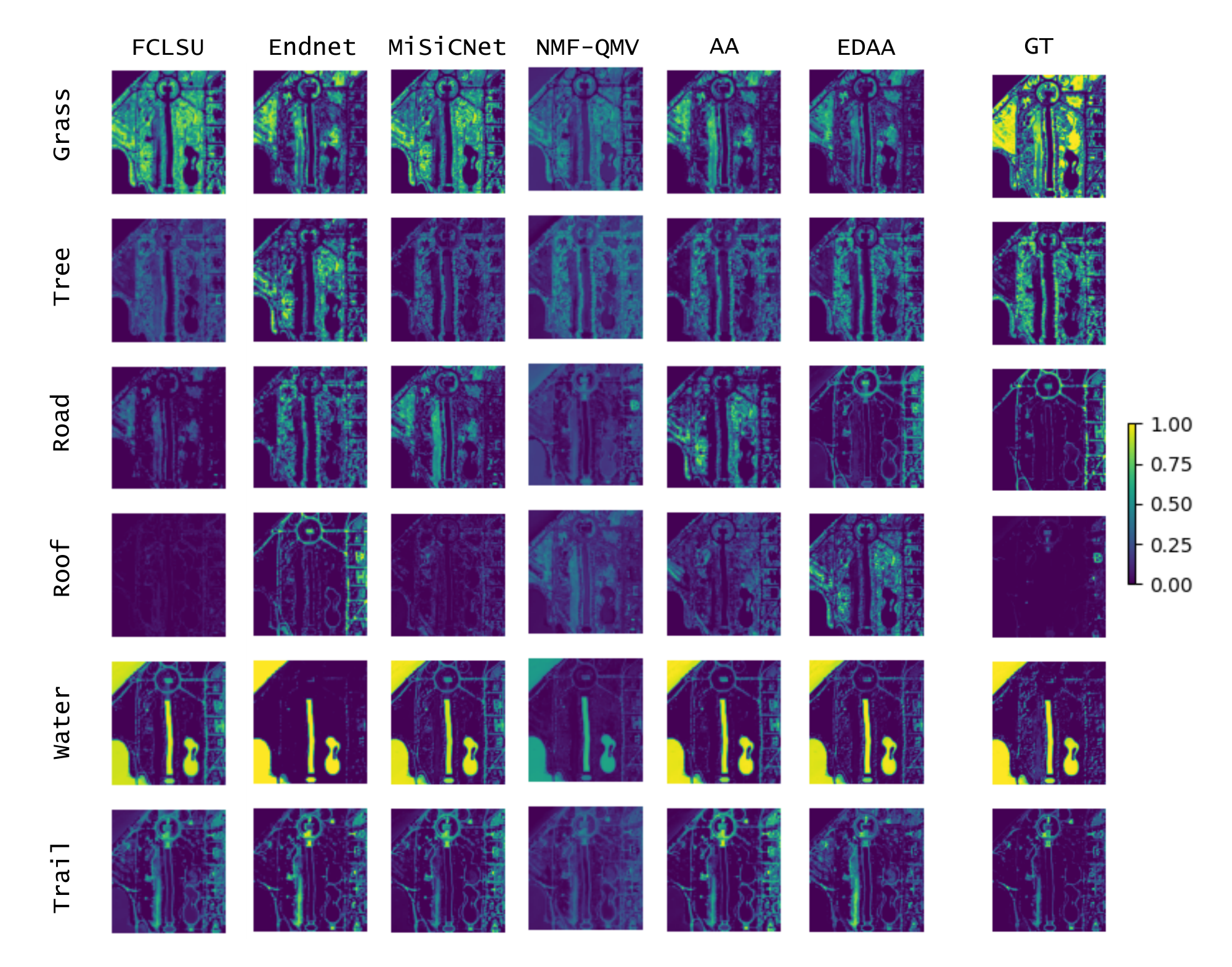}
  \caption{Estimated abundances on the Washington DC dataset. The ground truth is
    displayed on the right side.}
  \label{fig:WDCA}
\end{figure*}

Tables \ref{table:RMSE} and \ref{table:SAD} report the unmixing accuracy in
terms of abundances RMSE (\ref{eq:RMSE}) and endmembers SAD (\ref{eq:SAD}) on
six standard real datasets.
The datasets are arbitrarily ranked based on their difficulty.
For a fair comparison, all methods were evaluated on the $\ell_2$-normalized
data (\ref{eq:normalization}) which induces slight changes compared to the
results reported in \cite{rasti_misicnet_2022}.

Overall EDAA obtains the best abundances RMSE on all six datasets.
As the unmixing difficulty increases, the gap between EDAA and the other methods
grows larger.
In particular, plain AA is not able to tackle more complicated
mixing scenarios as in Urban6, APEX and WDC. 
We argue that our model selection technique is instrumental in avoiding
collapsing runs in which endmembers spectra are highly correlated.
This is underlined by the overall competitive SAD results obtained by EDAA
across datasets, especially on Urban6, APEX and WDC where Endnet is the only
contender.

The FCLSU baseline based on VCA obtains rather poor results except for WDC.
This is likely due to the pure pixel assumption. Indeed, VCA selects a single
pixel to represent the endmembers spectra, which is too stringent in real
scenarios where spectral variability is ubiquitous.

Despite its quadratic minimum volume \emph{boundary} term, NMF-QMV generally
obtains worse results than the FCLSU baseline.
Since it operates the unmixing in a subspace, NMF-QMV cannot prevent the endmembers spectra from having negative
values, which breaks the physical interpretability of the estimates and
subsequently harms the unmixing performance.
This phenomenon can be observed in figures
\ref{fig:JasperRidgeE},
\ref{fig:Urban6E}, \ref{fig:APEXE} and \ref{fig:WDCE} for
several endmembers.
The associated abundances in figures
\ref{fig:JasperRidgeA},
\ref{fig:Urban6A}, \ref{fig:APEXA} and \ref{fig:WDCA} show
that NMF-QMV produces maps that are too uniform and lack sparsity.

In contrast, Endnet achieves very good results in terms of SAD but tends to
create overly sparse abundances which hinders its performance in terms of
abundances RMSE.
For instance, as can be seen in figures \ref{fig:WDCE} and \ref{fig:WDCA}, the
Road endmember is overlooked by Endnet even though EDAA recovers it neatly.
Likewise, in figure \ref{fig:APEXA}, the Road endmember spreads too much
compared to EDAA which appears closer to the ground truth.

MiSiCNet gives better unmixing results than Endnet in terms of abundances RMSE
except for APEX although the SAD results falls in favor of Endnet except for
Jasper Ridge.
This is likely due to Endnet using the spectral angle distance on
the input data in its loss which helps in achieving better SAD accuracy.
However a good SAD is not sufficient to obtain good abundance maps, an area
where MiSiCNet tends to shine as it incorporates spatial information by using
convolutional filters and implicitly applying a regularizer on abundances.

Finally, AA is a very competitive method for the three simplest
datasets yet its performance decreases drastically when dealing with more complicated
mixtures.
For example, in figures \ref{fig:JasperRidgeE} and \ref{fig:JasperRidgeA} only
AA and EDAA are able to uncover the Road endmember in Jasper
Ridge whereas all the other techniques fail.
Yet, in figures \ref{fig:Urban6E} and \ref{fig:Urban6A}, AA completely misses the Metal and Dirt endmembers in Urban6
while EDAA correctly identify all endmembers and produce reasonable abundance maps.
Unlike EDAA, the random matrices initialization in AA cannot be leveraged to
create an appropriate model selection procedure due to its slowness.

Additional qualitative results for the Samson and Urban4 datasets can be found
in the supplementary material.

\subsection{Computational cost}

Table \ref{table:cost} reports the processing times for the different unmixing
algorithms on the six real datasets. NMF-QMV was implemented in Matlab (2020b) while
FCLSU, Endnet, MiSiCNet and the AA variants were implemented in Python (3.8).
NMF-QMV, FCLSU and AA run on CPU whereas Endnet, MiSiCNet and EDAA run on GPU.
The processing times were obtained using a computer with an Intel(R) Xeon(R)
Silver 4110 processor (2.10 GHz), 32 cores, 64GB of memory, and a NVIDIA
GeForce RTX (2080 Ti) graphical processing unit.
The table shows that FCLSU is clearly faster than the other unmixing techniques,
however it is a supervised method that relies on an endmembers extraction
algorithm. In this case, VCA is used which is also fast.
The deep learning methods are the slowest techniques despite running on GPU.
Interestingly, EDAA requires a lower computational cost than NMF-QMV and AA although our approach consists in aggregating 50 runs obtained
iteratively. 
For example, it takes on average 1.5 seconds for EDAA to perform a single unmixing
task on the Urban6 dataset, which is three times faster than FCLSU. 
This demonstrates the efficiency of EDAA which allows us to use an
adequate model selection procedure over several runs.

\begin{table}[h]
  \centering
  \captionof{table}{Processing times in seconds on six real datasets. The best
    results are in bold and the second best underlined. EDAA includes the model
    selection procedure over $M=50$ runs.}
  \begin{adjustbox}{width=0.5\textwidth}
    \begin{tabular}{c | c c c c c c}
  \toprule
  & FCLSU & Endnet & MiSiCNet & NMF-QMV & AA & EDAA \\
  \hline
  Samson & \textbf{0.3} & $\approx$ 560 & $\approx$ 80 & 20.1 & 17.6 & \underline{11.2} \\
  JasperRidge & \textbf{0.4} & $\approx$ 680 & $\approx$ 90 & 22.3 & 21.9 & \underline{9.6} \\
  Urban4 & \textbf{3.6} & $\approx$ 720 & $\approx$ 411 & 112.5 & 237.6 & \underline{66.3} \\
  Urban6 & \textbf{4.4} & $\approx$ 1000 & $\approx$ 417 & 158.4 & 343.0 & \underline{75.2} \\
  APEX & \textbf{0.6} & $\approx$ 720 & $\approx$ 92 & 27.2 & 44.8 & \underline{15.4} \\
  WDC & \textbf{4.0} & $\approx$ 1000 & $\approx$ 409 & 174.4 & 361.8 & \underline{81.0} \\
  \bottomrule
\end{tabular}

  \end{adjustbox}
  \label{table:cost}
\end{table}

\subsection{Ablation study} \label{subsec:ablation}

\begin{figure*}[h]
  \centering
  \subfloat[]{\includegraphics[width=0.33\textwidth]{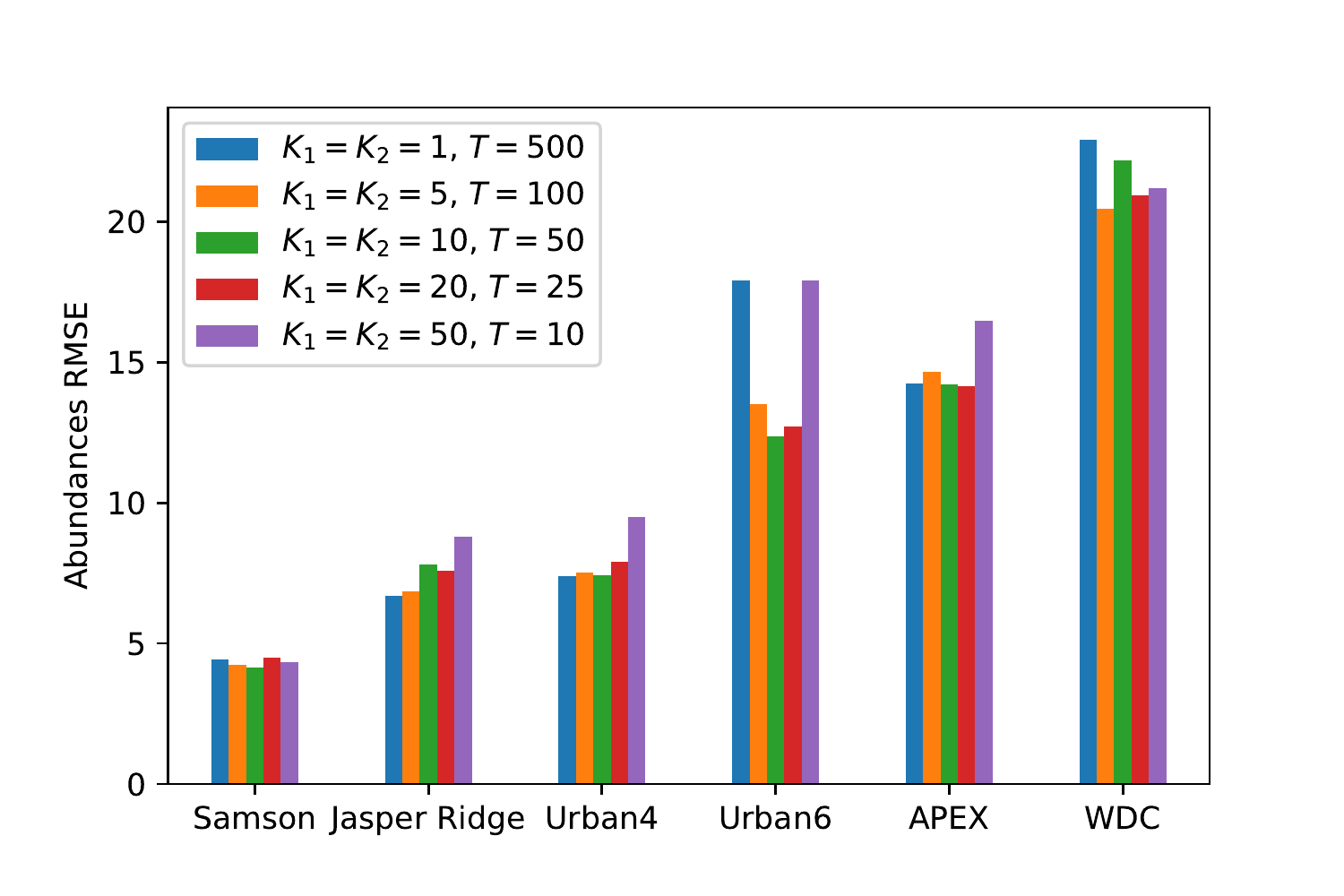}}
  \hfil
  \subfloat[]{\includegraphics[width=0.33\textwidth]{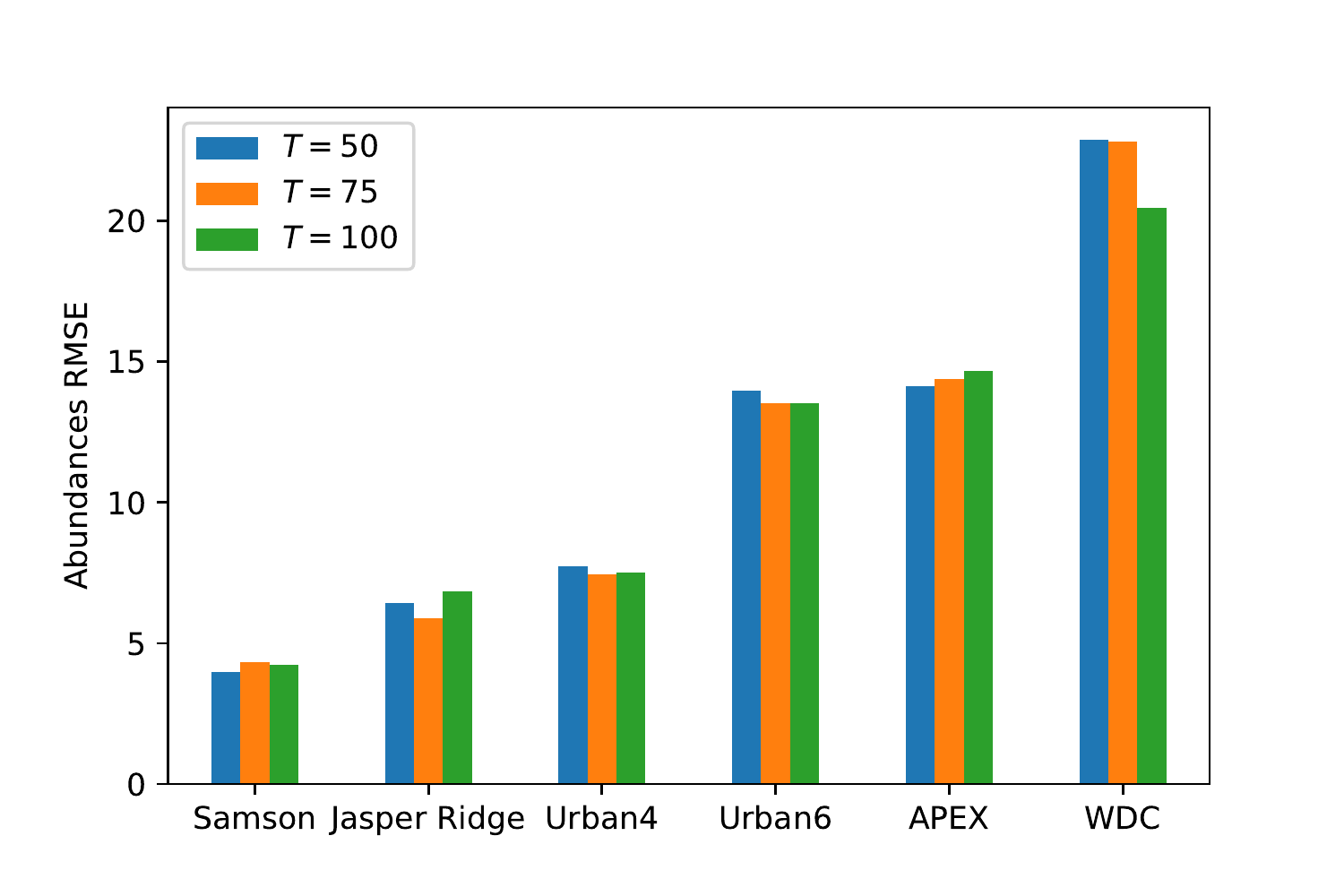}}
  \hfil
  \subfloat[]{\includegraphics[width=0.33\textwidth]{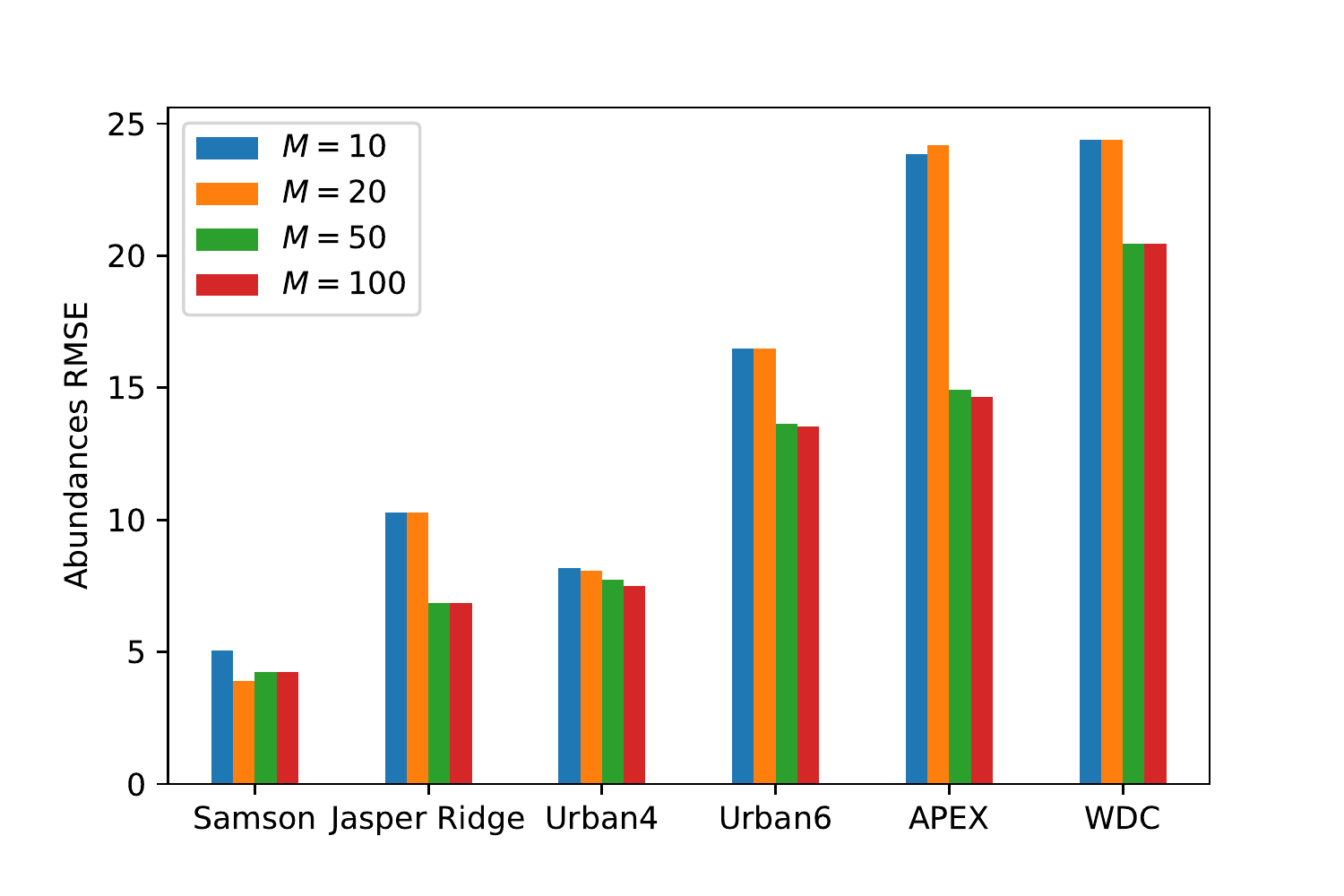}}
  \caption{Sensitivity analysis to the hyperparameters in Algorithms
    \ref{alg:EDAA} and \ref{alg:criterion} measured in global abundances RMSE: (a) Varying
  inner and outer iterations $K_1, K_2$ and $T$ for a constant number of updates
  (1000) and runs $M=100$, (b) Varying
outer iterations $T$ using $K_1=K_2=5$ and (c) Varying number of runs $M$ using
$K_1=K_2=5$ and $T=100$.}
  \label{fig:ablation}
\end{figure*}

Finally, we study the sensitivity to hyper-parameters for Algorithm \ref{alg:EDAA} and
\ref{alg:criterion} in figure \ref{fig:ablation} where the Y-axis corresponds to the
overall abundances RMSE.
Given a fixed computational budget of 1000 updates, figure \ref{fig:ablation}
(a) shows that the hyper-parameters of EDAA are robust provided that the number
of runs $M$ in the model selection is large enough (here 100). Only the two extremes
configurations ($K_1=K_2=1$, $T=500$ and $K_1=K_2=50$, $T=10$) are slightly
worse, especially on Urban6.
For the remaining experiments, we use $K_1=K_2=5$.
In figure \ref{fig:ablation} (b), we see that the number of outer iterations is
quite stable except for WDC which requires more updates (1000, \ie $T=100$).
Finally, we study the importance of the number of runs $M$ from which to select
the best candidate in figure \ref{fig:ablation} (c).
We observe that the model selection procedure requires at least 50 runs to
obtain very good performances, hence we use $M=50$ in our unmixing experiments.
On unknown datasets where real-time unmixing is not required, it is advised to
use a large number of runs (at least 100) to ensure that the model selection
procedure yields a good candidate.
Detailed results for both abundances RMSE and SAD metrics are available in the
supplementary material.

\section{Conclusion}

\label{sec:ccl}

We have proposed a new algorithm based on archetypal analysis for blind
hyperspectral unmixing.
We have shown how to take advantage of its efficient GPU implementation in order
to develop an adequate model selection procedure to obtain state-of-the-art
performances.
Remarkably, our simple and easy-to-use approach considerably improves the unmixing
results on a comprehensive collection of standard real datasets.
Finally, we have made our results reproducible by releasing an open source
codebase which also includes the plain archetypal analysis variant presented in this study.

\subsection*{Acknowledgments and Funding}

This project was supported by the ERC grant number 714381 (SOLARIS project) and
by ANR 3IA MIAI@Grenoble Alpes (ANR-19-P3IA-0003).

\bibliographystyle{IEEEtran}
\bibliography{hsu}

\clearpage

\begin{appendices}

\clearpage
\section{Detailed datasets description}

In this section, we provide illustrations of the unmixing datasets
used in the experiments. Each dataset is described with a false-color RGB image
alongside the $\ell_2$-normalized ground truth endmembers.

  \begin{figure}[h]
    \centering
    \subfloat[]{\includegraphics[height=1in]{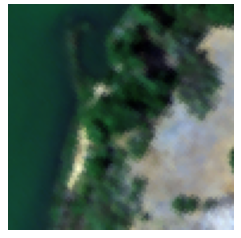}}
    \hfil
    \subfloat[]{\includegraphics[height=1in]{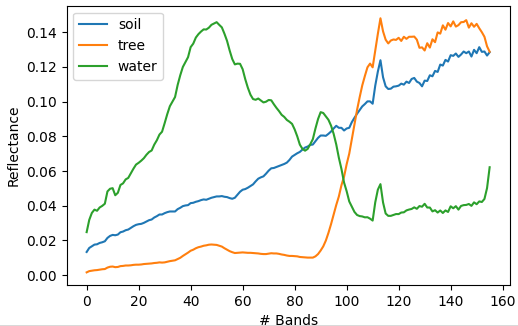}}
    \caption{Samson dataset: (a) False colors RGB image (Red: 83rd band, Green: 43, Blue: 9) (b)
    $\ell_2$-normalized ground truth endmembers.}
    \label{fig:samson}
  \end{figure}

  \begin{figure}[h]
    \centering
    \subfloat[]{\includegraphics[height=1in]{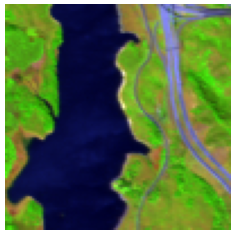}}
    \hfil
    \subfloat[]{\includegraphics[height=1in]{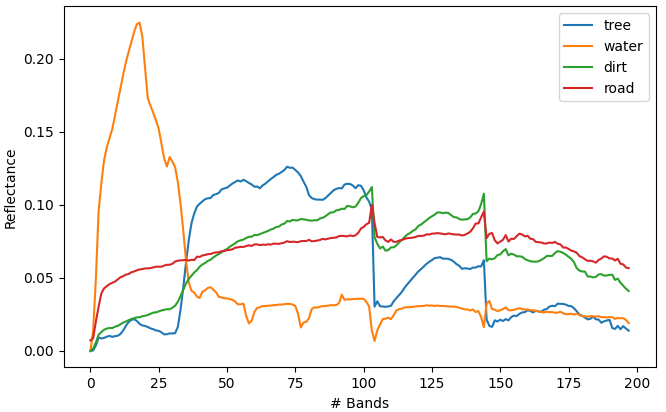}}
    \caption{JasperRidge dataset: (a) False colors RGB image (Red: 130th band, Green: 50, Blue:
    5) (b) $\ell_2$-normalized ground truth endmembers.}
    \label{fig:jasper}
  \end{figure}

  \begin{figure}[h]
    \centering
    \subfloat[]{\includegraphics[height=1in]{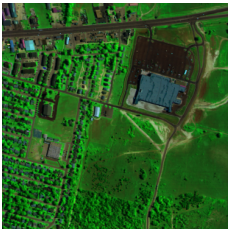}}
    \hfil
    \subfloat[]{\includegraphics[height=1in]{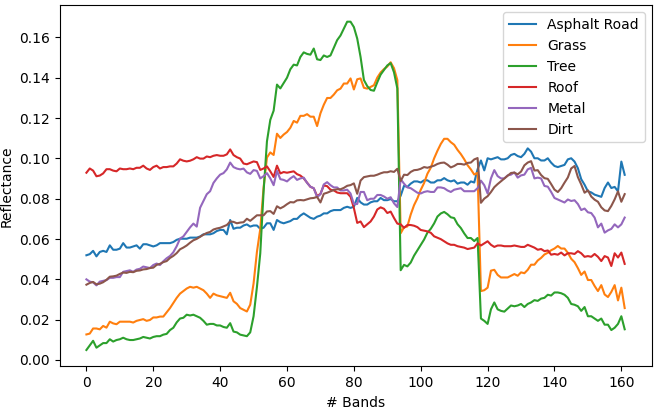}}
    \caption{Urban dataset: (a) False colors RGB image (Red: 130th band, Green: 70, Blue: 30) (b)
    $\ell_2$-normalized ground truth endmembers for Urban6. Urban4 corresponds to the first 4
    materials.}
    \label{fig:urban}
  \end{figure}

  \begin{figure}[h]
    \centering
    \subfloat[]{\includegraphics[height=1in]{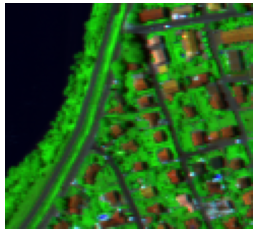}}
    \hfil
    \subfloat[]{\includegraphics[height=1in]{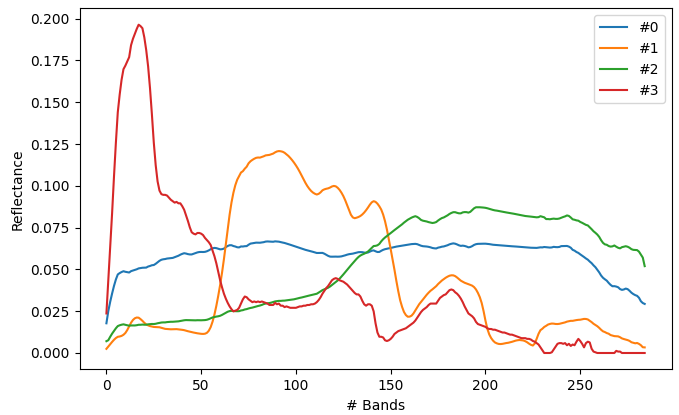}}
    \caption{APEX dataset: (a) False colors RGB image (Red: 200th band, Green: 100, Blue: 10) (b)
    $\ell_2$-normalized ground truth endmembers (\#0: Road, \#1: Tree, \#2: Roof,
    \#3: Water).}
    \label{fig:apex}
  \end{figure}

  \begin{figure}[h]
    \centering
    \subfloat[]{\includegraphics[height=1in]{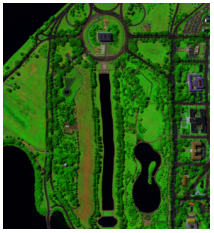}}
    \hfil
    \subfloat[]{\includegraphics[height=1in]{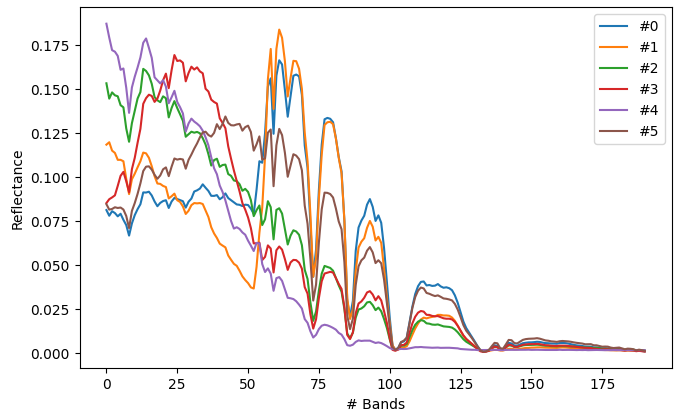}}
    \caption{Washington DC Mall dataset: (a) False colors RGB image (Red: 150th band, Green: 75,
    Blue: 20) (b) $\ell_2$-normalized ground truth endmembers (\#0: Grass, \#1:
    Tree, \#2: Road, \#3: Roof, \#4: Water, \#5: Trail).}
    \label{fig:wdc}
  \end{figure}

\section{Additional  results}

\paragraph{Unmixing experiments}

\begin{figure*}[h]
  \centering
  \includegraphics[width=\textwidth]{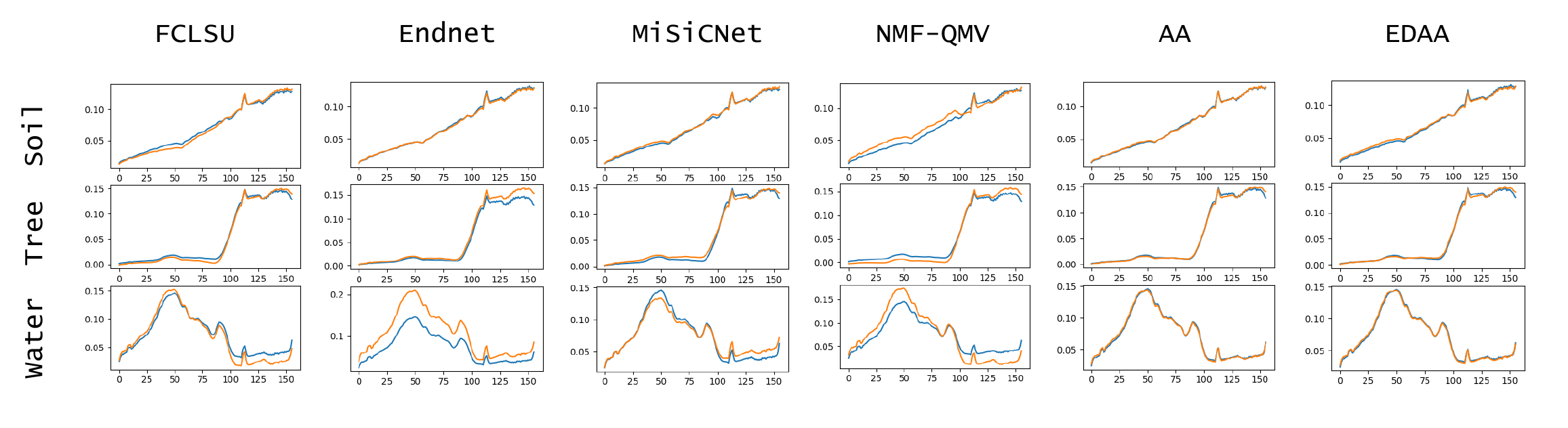}
  \caption{Estimated endmembers on the Samson dataset.
    Ground truth endmembers are displayed in blue while their estimates are in orange.}
  \label{fig:SamsonE}
\end{figure*}

\begin{figure*}[h]
  \centering
  \includegraphics[width=\textwidth]{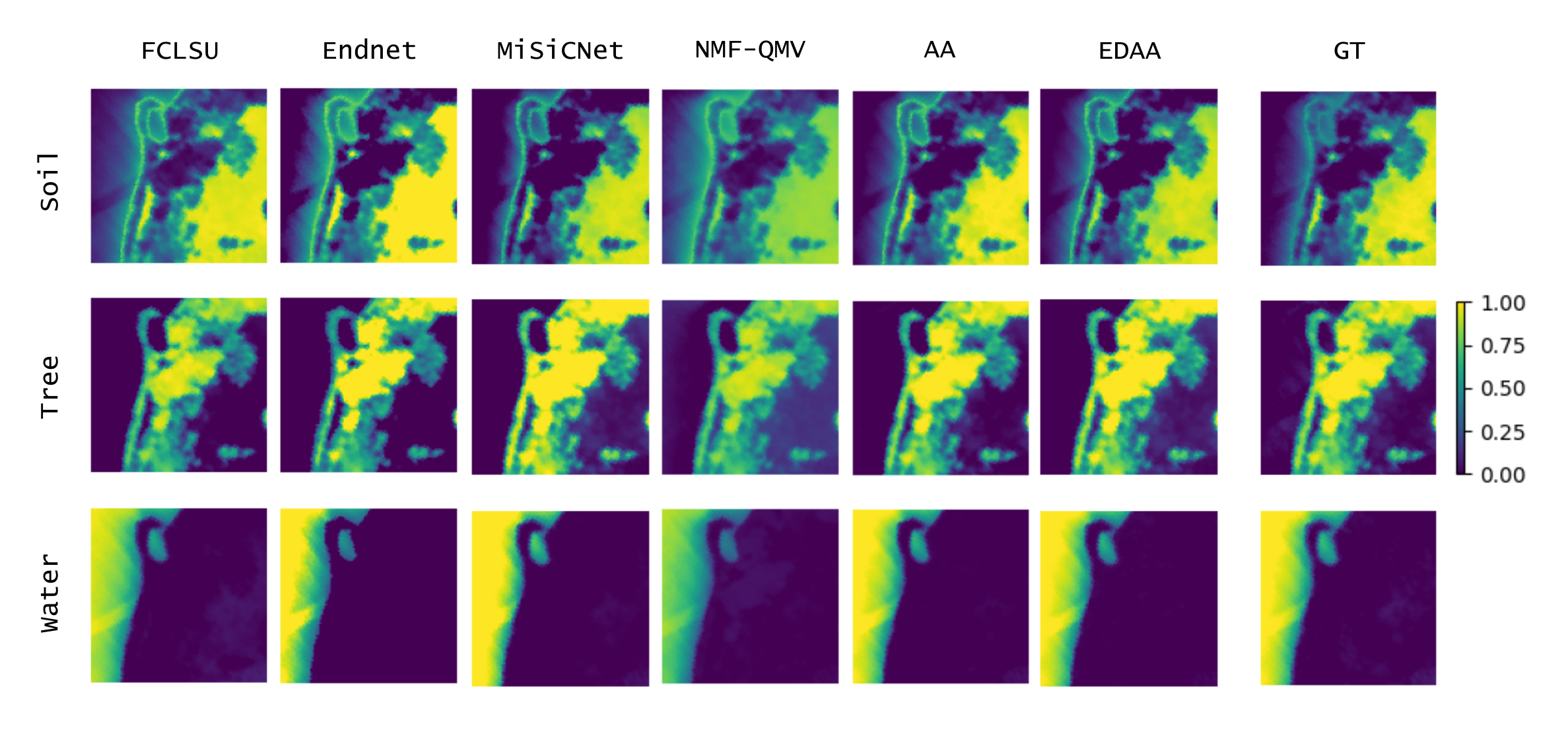}
  \caption{Estimated abundances on the Samson dataset. The ground truth is
  displayed on the right side.}
  \label{fig:SamsonA}
\end{figure*}

\begin{figure*}[h]
  \centering
  \includegraphics[width=\textwidth]{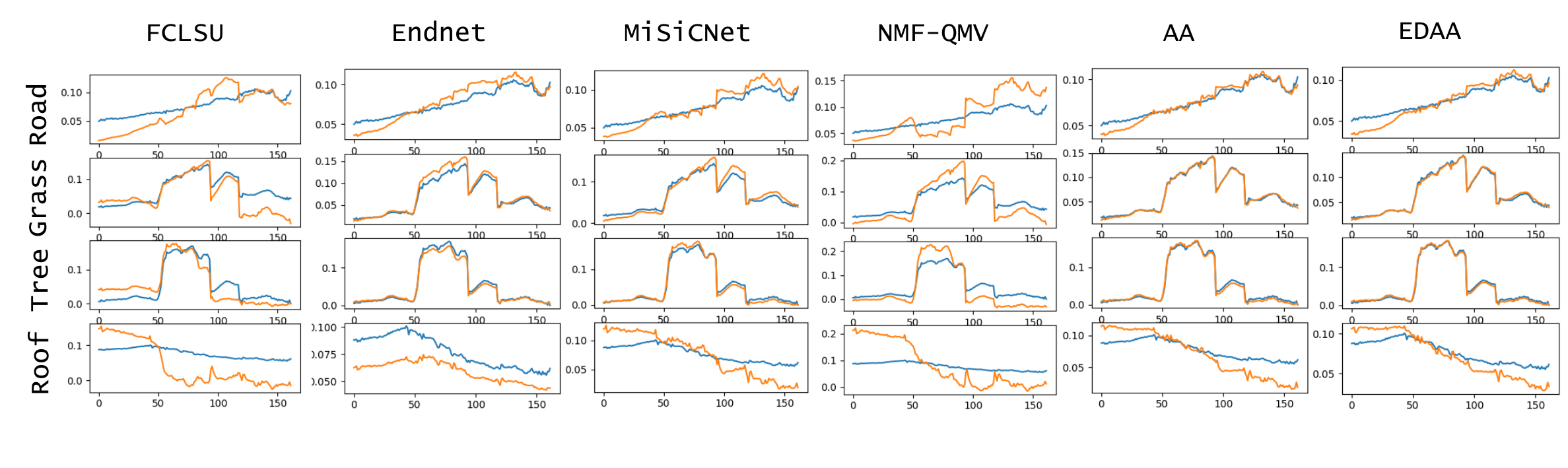}
  \caption{Estimated endmembers on the Urban4 dataset.
    Ground truth endmembers are displayed in blue while their estimates are in orange.
  }
  \label{fig:Urban4E}
\end{figure*}

\begin{figure*}[h]
  \centering
  \includegraphics[width=\textwidth]{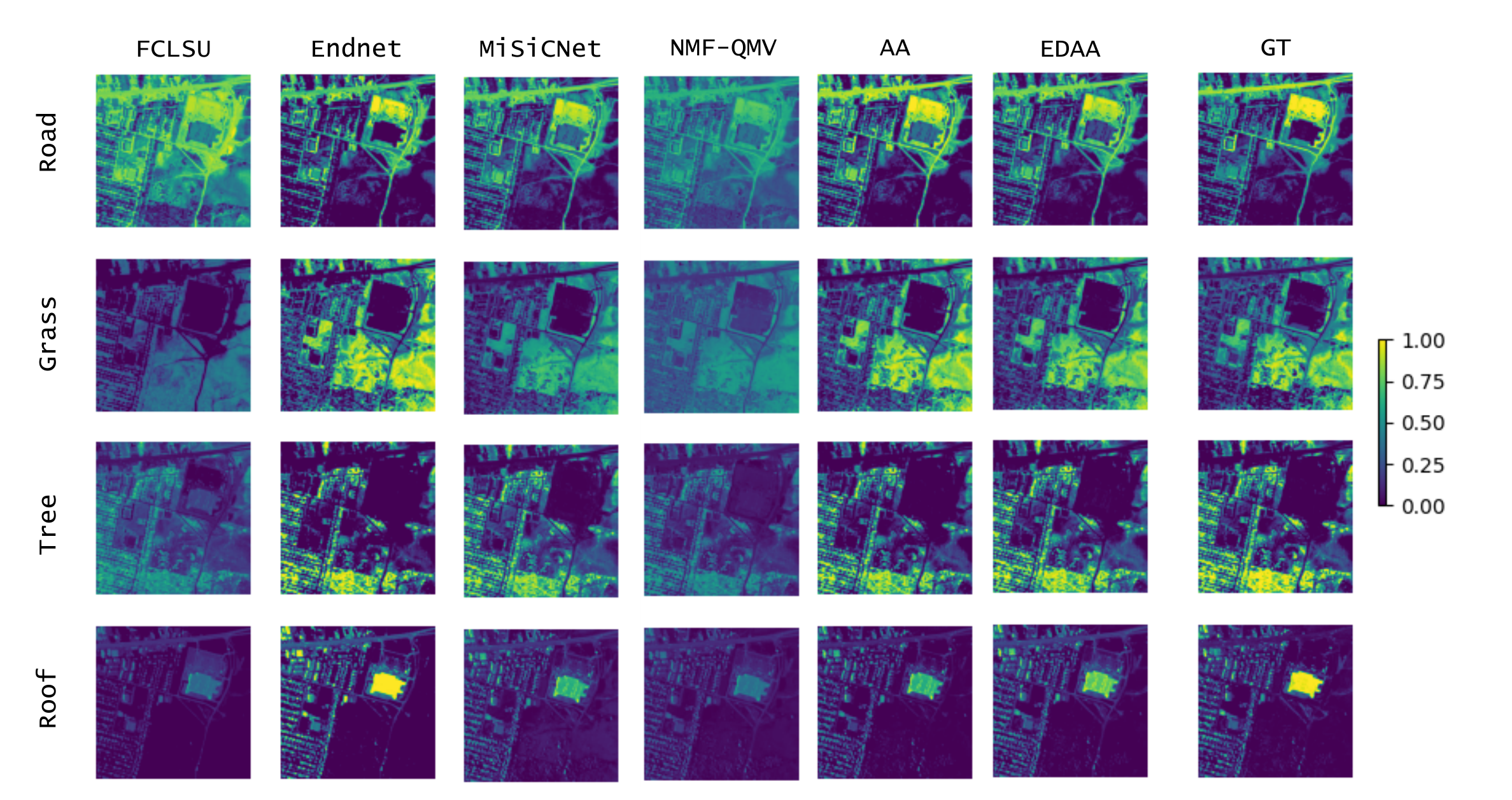}
  \caption{Estimated abundances on the Urban4 dataset. The ground truth is
  displayed on the right side.}
  \label{fig:Urban4A}
\end{figure*}

We provide qualitative results on the Samson and Urban4 datasets in figures
\ref{fig:SamsonE}, \ref{fig:SamsonA}, \ref{fig:Urban4E} and \ref{fig:Urban4A}.

\paragraph{Ablation study}

We report the detailed results obtained in the ablation study.
For each dataset, the overall abundances RMSE and SAD are computed for all configurations.
Table \ref{table:alg1_constant} studies the sensitivity of the inner and outer
iterations $K_1$, $K_2$ and $T$ in EDAA given a fixed computational budget.
Table \ref{table:alg1_decrease} studies the sensitivity of the outer iterations $T$ in EDAA when we decrease the computational budget.
Finally, table \ref{table:alg2} underlines the importance of the number of runs
$M$ in the model selection procedure.

\begin{table*}[h]
  \centering
  \captionof{table}{Sensitivity to hyperparameters of EDAA
    for a constant number of updates (1000). The abundances RMSE and SAD metrics are computed globally. The best results are in bold and the second best are underlined.}
  \begin{adjustbox}{width=\textwidth}
    \begin{tabular}{c | c | c c c c c}
  \toprule
  & & $K_1=K_2=1$, $T=500$ & $K_1=K_2=5$, $T=100$  & $K_1=K_2=10$, $T=50$ & $K_1=K_2=20$, $T=25$ & $K_1=K_2=50$, $T=10$ \\
  \hline
  \multirow{2}{*}{Samson}
  & RMSE & 4.42 & \underline{4.24} & \textbf{4.15} & 4.48 & 4.34 \\
  & SAD & 1.65 & \underline{1.64} & \textbf{1.61} & 1.78 & 1.68 \\
  \hline
  \multirow{2}{*}{Jasper Ridge}
  & RMSE & \textbf{6.70} & \underline{6.85} & 7.80 & 7.59 & 8.79 \\
  & SAD & 3.48 & \underline{3.22} & 4.28 & \textbf{3.12} & 4.79 \\
  \hline
  \multirow{2}{*}{Urban4}
  & RMSE & \textbf{7.40} & 7.51 & \underline{7.43} & 7.90 & 9.49 \\
  & SAD & 5.72 & 5.87 & \underline{5.53} & 6.05 & \textbf{5.46} \\
  \hline
  \multirow{2}{*}{Urban6}
  & RMSE & 17.92 & 13.52 & \textbf{12.35} & \underline{12.71} & 17.92 \\
  & SAD & 11.71 & \textbf{7.95} & 8.74 & \underline{8.52} & 8.79 \\
  \hline
  \multirow{2}{*}{APEX}
  & RMSE & 14.25 & 14.66 & \underline{14.20} & \textbf{14.15} & 16.46\\
  & SAD & 8.17 & \underline{6.29} & 7.73 & 8.06 & \textbf{5.26} \\
  \hline
  \multirow{2}{*}{WDC}
  & RMSE & 22.91 & \textbf{20.47} & 22.16 & \underline{20.92} & 21.20 \\
  & SAD & \underline{12.59} & \textbf{12.33} & 16.05 & 13.46 & 13.51 \\
  \hline

  \bottomrule
\end{tabular}
  \end{adjustbox}
  \label{table:alg1_constant}
\end{table*}

\begin{table}[h]
  \centering
  \captionof{table}{Sensitivity to the number of outer iterations $T$ of Algorithm
    EDAA with $K_1=K_2=5$. The abundances RMSE and SAD metrics are computed globally. The best results are in bold and the second best are underlined.}
    \begin{tabular}{c | c | c c c}
  \toprule
  & & $T=100$  & $T=75$ & $T=50$ \\
  \hline
  \multirow{2}{*}{Samson}
  & RMSE & \underline{4.24} & 4.34 & \textbf{3.97} \\
  & SAD &  \underline{1.64} & 1.69 & \textbf{1.46} \\
  \hline
  \multirow{2}{*}{Jasper Ridge}
  & RMSE &  6.85 & \textbf{5.90} & \underline{6.44} \\
  & SAD &  3.22 & \textbf{3.06} & \underline{3.19} \\
  \hline
  \multirow{2}{*}{Urban4}
  & RMSE & \underline{7.51} & \textbf{7.46} & 7.72 \\
  & SAD & 5.87 & \underline{5.01} & \textbf{4.67} \\
  \hline
  \multirow{2}{*}{Urban6}
  & RMSE & \textbf{13.52} & \underline{13.54} & 13.99 \\
  & SAD &  7.95 & \textbf{7.85} & \underline{7.93} \\
  \hline
  \multirow{2}{*}{APEX}
  & RMSE & 14.66 & \underline{14.38} & \textbf{14.12} \\
  & SAD & \textbf{6.29} & \underline{6.58} & 6.99 \\
  \hline
  \multirow{2}{*}{WDC}
  & RMSE & \textbf{20.47} & \underline{22.81} & 22.89 \\
  & SAD & \textbf{12.33} & 16.97 & \underline{16.87} \\
  \hline

  \bottomrule
\end{tabular}
  \label{table:alg1_decrease}
\end{table}

\begin{table}[h]
  \centering
  \captionof{table}{Sensitivity to the number of runs $M$ of the model selection
    procedure with $K_1=K_2=5$ and $T=100$. The abundances RMSE and SAD metrics are computed globally. The best results are in bold and the second best are underlined.}
    \begin{tabular}{c | c | c c c c}
  \toprule
  & & $M=10$  & $M=20$ & $M=50$ & $M=100$ \\
  \hline
  \multirow{2}{*}{Samson}
  & RMSE & 5.05 & \textbf{3.90} & \underline{4.24} & \underline{4.24} \\
  & SAD & 1.94 & \textbf{1.43} & \underline{1.64} &\underline{1.64} \\
  \hline
  \multirow{2}{*}{Jasper Ridge}
  & RMSE & 10.27 & 10.27 & \textbf{6.85} & \textbf{6.85} \\
  & SAD & 3.73 & 3.73 & \textbf{3.22} & \textbf{3.22} \\
  \hline
  \multirow{2}{*}{Urban4}
  & RMSE & 8.17 & 8.06 & \underline{7.75} & \textbf{7.51} \\
  & SAD & 5.43 & \underline{5.41} & \textbf{5.36} & 5.87 \\
  \hline
  \multirow{2}{*}{Urban6}
  & RMSE & 16.50 & 16.50 & \underline{13.63} &\textbf{13.52} \\
  & SAD & 8.16 & 8.16 & \textbf{7.92} & \underline{7.95} \\
  \hline
  \multirow{2}{*}{APEX}
  & RMSE & 23.84 & 24.17 & \underline{14.94} & \textbf{14.66} \\
  & SAD & 12.57 & 12.88 & \textbf{6.06} &\underline{6.29} \\
  \hline
  \multirow{2}{*}{WDC}
  & RMSE & 24.39 & 24.39 & \textbf{20.46} &\underline{20.47} \\
  & SAD & \textbf{11.52} & \textbf{11.52} & 12.40 & 12.33 \\
  \hline

  \bottomrule
\end{tabular}
  \label{table:alg2}
\end{table}

\end{appendices}

\end{document}